\documentclass[%
 reprint,
superscriptaddress,
 amsmath,amssymb,
 aps,
]{revtex4-2}

\usepackage{graphicx}
\usepackage{dcolumn}
\usepackage{bm}
\usepackage{hyperref}
\usepackage{soul}
\usepackage{xcolor}

\DeclareMathOperator\erf{erf}

\makeatother

\begin{document}

\preprint{APS/123-QED}

\title{Crystal Nucleation Rates in One-Component Yukawa Systems}

\author{B.~Arnold}
\affiliation{Laboratory for Laser Energetics, University of Rochester, Rochester, New York 14623, USA}
\affiliation{Department of Physics and Astronomy, University of Rochester, Rochester, New York 14611, USA }

\author{J.~Daligault}
\author{D.~Saumon}
\affiliation{Los Alamos National Laboratory, P.O. Box 1663, Los Alamos, New Mexico 87545, USA}

\author{Antoine B\'edard}
\affiliation{Department of Physics, University of Warwick, Coventry CV4 7AL, UK }

\author{S. X.~Hu}
\email{corresponding author: shu@lle.rochester.edu}
\affiliation{Laboratory for Laser Energetics, University of Rochester, Rochester, New York 14623, USA}
\affiliation{Department of Physics and Astronomy, University of Rochester, Rochester, New York 14611, USA }
\affiliation{Department of Mechanical Engineering, University of Rochester, Rochester, New York 14611, USA }

\date{\today}

\begin{abstract}
Nucleation in the supercooled Yukawa system is relevant for addressing current challenges in understanding a range of crystallizing systems including white dwarf (WD) stars. We use both brute force and seeded molecular dynamics simulations to study homogeneous nucleation of crystals from supercooled Yukawa liquids. With our improved approach to seeded simulations, we obtain quantitative predictions of the crystal nucleation rate and cluster size distributions as a function of temperature and screening length. These quantitative results show trends towards fast nucleation with short-ranged potentials. They also indicate that for temperatures $T>0.9T_m$, where $T_m$ is the melt temperature, classical homogeneous nucleation is too slow to initiate crystallization but transient clusters of $\sim 100$ particles should be common. We apply these general results to a typical WD model and obtain a delay of  $\sim 0.6$ Gyr in the onset of crystallization that may be observable.
\end{abstract}

\maketitle

\section{Introduction}

Understanding nucleation of crystals from supercooled liquids is critical for physical systems in fields ranging from materials manufacturing to planetary science \cite{BC8, ruiz2017, Huguet2018, wilson2023, Nederstigt2023, wang2023}. Even white dwarf stars have been observed to crystallize \cite{Tremblay2019}. Fully explaining observations of WD populations will require improved understanding of  crystallization in dense plasmas \cite{blouinCOphasediagram2020}. Despite the importance of nucleation across many fields, classical nucleation theory (CNT), the dominant framework describing nucleation, fails to quantitatively predict nucleation rates in many systems. By calculating rates of nucleation in dense plasmas, we elucidate aspects of nucleation that apply to all systems that undergo crystal nucleation, particularly focusing on the relevance to WD.

The lack of a predictive theory of nucleation in general leads to a situation where every system of interest must be studied individually \cite{jChemPhysNucleaitonSpecialIssue}. Most of these nucleation studies focus on single-component model systems like hard spheres \cite{volkov2002, Auer2004, Gispen2023} or Lennard-Jones particles \cite{Honeycutt1987, tenWolde1996, seededLJ}. They have revealed some mechanisms behind nucleation and the qualitative effect of temperature on the process, but there are still many unanswered questions. There is little understanding of how, for instance, nucleation in mixtures differs from single-component systems or how the form of interatomic interactions influences the process. 

We focus on the Yukawa one-component plasma (YOCP), a collection of classical particles interacting via the screened Coulomb potential 
\begin{equation}
    V(r) = \frac{Q^2}{r} e^{-\kappa r/a}
\end{equation}
where $a= \left( \frac{3}{4 \pi n} \right)^{1/3}$ is the average interparticle distance and $n$ is the ion number density. $\kappa$ is the unitless parameter that characterizes the electronic screening length relative to $a$. At thermal equilibrium at temperature $T$, the system is fully characterized by $\kappa$ and either the Coulomb coupling parameter $\Gamma = \frac{Q^2}{a k_B T}$ or reduced temperature $\Theta = \frac{T}{T_m}$ (where $T_m$ is the melt temperature). In the special case where $\kappa=0$ (the weak screening limit) it recovers the well-known One-Component Plasma (OCP) model. 

This pair potential is applicable for a range of physical systems including ions at the extreme densities inside white dwarf (WD) stars ($>10^6 \text{g}/\text{cm}^3$). At these conditions, atoms are completely pressure ionized and a background of partially relativistic degenerate electrons provides screening characterized by $\kappa \approx 0.3$. As WD slowly cool over billions of years, this plasma crystallizes \cite{vanHorn1968}. During this process, species of different masses gravitationally segregate and crystallization releases latent heat, which slows the star's cooling \cite{Tremblay2019}. Observed anomalies in the cooling delays of certain WD \cite{Qbranch} cannot be explained without an improved understanding of the formation of solid clusters around freezing \cite{blouinNeDist2021, Bedard2024}. 

While the primary relevance of this work lies in understanding the fundamental mechanisms of crystal nucleation in dense astrophysical plasmas, it is worth noting the broader connection to studies on crystal formation in systems of strongly interacting electric charges, such as dusty plasmas and charged colloids. In these systems, interactions between charged particles give rise to similarly remarkable phenomena of self-organization and crystallization \cite{ISScomplexplasma}. Insights from our study contribute to a deeper understanding of these related systems, particularly in shedding light on how long-range electrostatic forces drive phase transitions across diverse physical contexts. Conversely, studies of dusty plasmas, with their well-controlled experimental conditions, could also provide valuable insights into the complex processes underlying crystal nucleation in dense Coulomb systems.

Previously, nucleation of the OCP and YOCP has been studied analytically \cite{Cooper2008} and with molecular dynamics (MD) simulation at fixed $\Theta$ \cite{daligault2006} and under rapid quenches \cite{Ogata1992, hammerberg1994}, but none of this work included nucleation rates at the weak undercooling $(\Theta>0.72)$ that is relevant to slowly cooling systems like WD.

\begin{figure}
    \centering
    \includegraphics[width=1\linewidth]{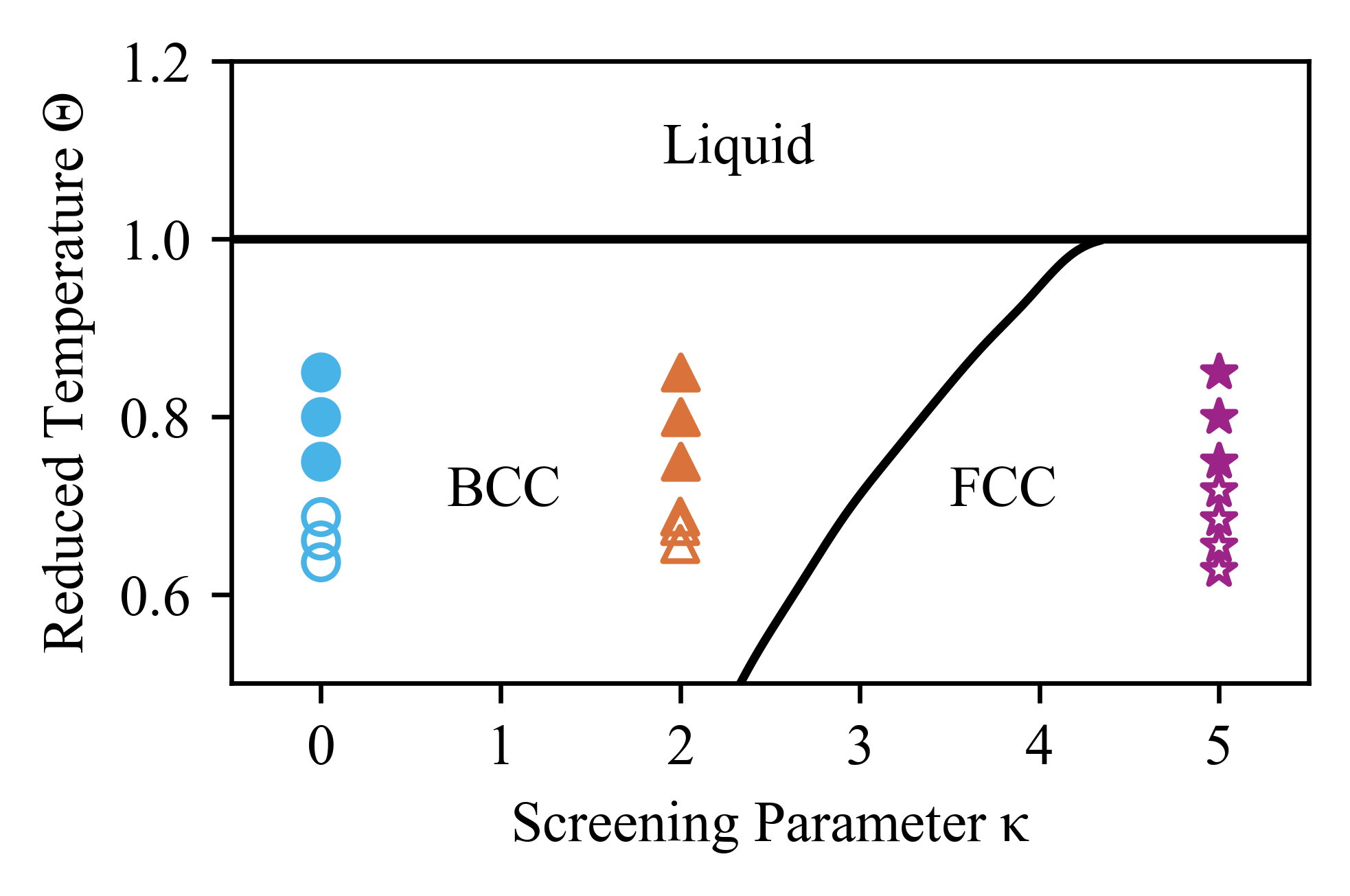}
    \caption{Phase diagram of the YOCP calculated in Ref. \protect{\cite{HamaguchiTriplePoint1997}}. At high temperatures the YOCP is always a liquid, but the low temperature phase transitions from BCC at weak screening (the OCP limit) to FCC in the strongly screened regime. Points represent the $(\kappa, \Theta)$ conditions at which we calculated nucleation rates (points here correspond to those in Fig. \protect{\ref{fig:nucRate}}). Open symbols represent low-temperature brute force simulations while filled symbols represent seeded simulations performed at higher temperatures.}
    \label{fig:phase_diagram}
\end{figure}

In this paper we extend previous work on YOCP nucleation in two ways. First, we employ both brute-force and seeded classical MD simulations to study homogeneous nucleation in a wider temperature range than is possible with brute force MD simulations alone. Second, we systematically study the impact of the range of interatomic interactions on nucleation by simulating $\kappa = 0,2,$ and $5$ Yukawa plasmas. Fig. \ref{fig:phase_diagram}
shows all $(\kappa, \Theta)$ points where we calculated homogeneous nucleation rates  (the number of macroscopic solid clusters formed per unit time per particle). This primary result is important because a system will not solidify unless the nucleation rate is sufficiently high for at least one nucleation event to occur over a relevant timescale.

We also use CNT to interpret our simulation results and predict the sizes of transient clusters present in the supercooled bulk liquid before freezing. Calculations like this will be useful when extending this work to WD-relevant multi-component mixtures because clusters with compositions distinct from the liquid are thought to enhance transport of heavy elements towards the core of the star more efficiently than single particle diffusion \cite{blouinNeDist2021}, but only if the clusters are sufficiently large \cite{Bauer2020}.

This paper is organized as follows: section \ref{sec:methods} discusses classical nucleation theory and the variety of molecular methods that could be used to study cluster formation. Sections \ref{sec:bruteForce} and \ref{sec:seededSimulation} explain the simulations and analysis we performed with brute force and seeded molecular dynamics, respectively. The analyses in section \ref{sec:seededSimulation} include several methodological improvements to previous seeded simulation studies, reducing some sources of error in nucleation rate estimates. We combine the results from these simulations to present nucleation rates and cluster size distributions for the YOCP with an application to WD cooling in section \ref{sec:results}. 

\section{Methods for Studying Nucleation}\label{sec:methods}

When a pure liquid cools below its freezing temperature it does not immediately freeze because the liquid is metastable for mild undercoolings. The traditional picture is that small crystal nuclei continually form and dissolve within the liquid until their size exceeds a critical value beyond which the solid phase will grow. This process of initiating the liquid-solid transition is known as nucleation. We consider only homogeneous nucleation, meaning that nuclei form only in the bulk. Nucleation is difficult to study because there are multiple possible nucleation mechanisms and the timescale of nucleation is in general much longer than timescales that can be easily simulated. 

When studying nucleation, the typical goals are to calculate the nucleation rate and identify the nucleation mechanism. The nucleation rate is simply understood as the number of nucleation events (thermal fluctuations leading to a region of the supercooled liquid solidifying) that occur per unit time, per unit volume. This constrains the rate at which a macroscopic sample can solidify. The nucleation mechanism is the pathway that the system traverses to get from the bulk liquid to a solid cluster. We allow only classical nucleation in our calculations of the nucleation rate, meaning that the pathway is restricted to be consistent with the CNT pathway discussed below.

In the remainder of this section, we discuss CNT, the dominant framework that is used with varying degrees of success to describe nucleation experiments. We then briefly cover how molecular simulations can quantitatively supplement CNT and provide insight into the nucleation mechanism.

\subsection{Classical Nucleation Theory}

Classical nucleation theory is a phenomenological theory that describes the kinetic process by which nucleation takes place. A first-principles analytical approach to this problem is intractable, so CNT makes progress by introducing several assumptions. In its simplest form, it assumes that nucleation is homogeneous and single-step, and that nuclei are isolated, which allows one to consider a single cluster of the solid phase forming from the undercooled liquid. The goal is to calculate the nucleation rate by assessing the probability per unit time per unit volume of such a cluster becoming macroscopic. This section gives a brief overview of how CNT arrives at a nucleation rate. For a more detailed derivation and some extensions of the CNT expressions presented here, see Ref. \cite{Kelton1991}.

CNT assumes that a solid cluster containing $N_s$ particles is spherical, has a infinitesimally thin interface with the surrounding liquid, and has a Gibbs free energy of formation, $\Delta G(N_s)$, given by
\begin{equation}
    \label{freeEnergy}\Delta G(N_s) = \Delta \mu N_s  + \gamma N_s^{2/3}.
\end{equation}
The first term ensures that the total free energy decreases by $\Delta \mu = \mu_s - \mu_\ell$ (the chemical potential difference between the bulk solid and bulk liquid at constant pressure) when a particle joins the solid cluster because the solid is thermodynamically favorable over the liquid. The second term is the unfavorable $(\gamma>0)$ surface free energy cost of creating a liquid-solid interface. Small clusters have large surface area to volume ratios, so the free energy penalty associated with the surface creates a ``barrier'' in $\Delta G$ making their growth unfavorable. Setting $\frac{d}{dN} \Delta G(N) = 0$ shows that the top of this barrier occurs at 
\begin{equation}
    \label{critSize}N^* = \left(\frac{2 \gamma}{3 \left| \Delta \mu\right| }\right)^3.
\end{equation}
For clusters larger than this critical size $N^*$, growth is favorable. 

In CNT, the undercooled liquid is composed of an ensemble of solid clusters whose sizes grow or diminish by the addition or loss of single particles. The rate of change of $P(N_S)$, the probability of finding a cluster of size $N_S$, follows a master equation in terms of transition probabilities describing the change in $P(N_S)$ due to either the attachment or removal of a particle occurring at rates $D_{N_S}^+$ and $D_{N_S}^-$, respectively:
\begin{equation}
    \label{eq:masterEq}
    \frac{d P(N_S)}{dt} = J_{N_S-1}-J_{N_s}
\end{equation}
where $J_{N_s} = D_{N_S}^+ P(N_S)- D_{N_S+1}^- P(N_S+1)$ is the current of clusters passing over the size $N$.

To estimate the nucleation rate, one imagines creating a stationary non-equilibrium state in the system by destroying the nuclei that reach a size $N_{\text{max}}$ somewhat larger than the critical size $N^*$ and by returning the constituent atoms to the $N_s=1$ state of individual atoms. Under the steady-state conditions,
\begin{equation}
    \label{eq:preliminaryNuc_Rate}
    J = J_{N_S-1} = J_{N_S}
\end{equation}
corresponds to the desired nucleation rate. One then assumes that the rates $D_{N_S}^+$ and $D_{N_S}^-$ retain their equilibrium values out of equilibrium. At equilibrium, according to detailed balance, these rates are related by
\begin{equation}
    \label{eq:detailedBalance}
    P^{eq}(N_S)D_{N_S}^{eq,+} = P^{eq}(N_S+1)D_{N_S+1}^{eq,-},
\end{equation}
where $P^{eq}(N_S)$ is the probability per particle to find a cluster of size $N_S$ in equilibrium. This can be regarded as the probability that a spontaneous fluctuation results in the formation of a cluster of size $N_S$, which depends on the free energy $\Delta G(N_S)$ to form such a nucleus as 
\begin{equation}
\label{distEq}
    P^e(N_s) \propto \exp{\left\{-\Delta G(N_s)/k_BT\right\}}.
\end{equation}
Note that for small cluster sizes $N_s<N^*$, this cluster size distribution is valid even for the steady-state nonequilibrium distribution because the timescale for reaching the top of the free energy barrier is much longer than the equilibration time within the free energy well when $\exp{\left\{-\Delta G(N_s)\right\}}>>\exp{\left\{-\Delta G(N^*)\right\}}$. Thus we can use Eq. (\ref{distEq}) to estimate the distribution of subcritical clusters.

Combining Eqs. \ref{eq:preliminaryNuc_Rate}, \ref{eq:detailedBalance} and \ref{distEq}, one arrives at the nucleation rate per particle
\begin{equation}
    \label{nucRate}
    J = D_+ \sqrt{\frac{\left| \Delta \mu \right|}{6 \pi k_B T N^*}} \exp{\left\{-\Delta G\left(N^*\right)/k_BT\right\}}
\end{equation}
where $D_+$ is a characteristic rate of particles attaching to a critical cluster. The square root factor is called the Zeldovitch factor. It contains all relevant information about the difference between the equilibrium and nonequilibrium pictures of nucleation. In order to make use of this, the required parameters are $N^*$, $\Delta \mu$, $\gamma$, and $D_+$.   

Cooper and Bildsten studied nucleation in the OCP with CNT. They proposed a modification to the theory that incorporates an entropy of mixing between clusters of different sizes \cite{Cooper2008}. This leads to a different nucleation rate
which is many orders of magnitude too small to be consistent with our simulations. Therefore we use the standard expression given in Eq. (\ref{nucRate}).

\subsection{Molecular Dynamics Methods}

Molecular dynamics simulations are a popular way to assess the nucleation pathway (including checking the validity of CNT), calculate CNT parameters, and estimate nucleation rates for benchmarking other calculations.

Brute force simulations are the best way to check whether a system follows the CNT pathway and to benchmark nucleation rates at low temperatures. In a brute force simulation, a supercooled liquid is initialized and allowed to evolve in an unbiased way, simply following Newton's laws until solidification. Since this method allows any transition pathway, it reveals whether the pathway assumed by CNT is applicable to the simulated systems. These simulations are very accurate; they are limited mostly by the quality of the interatomic potentials, thermostats interfering with the exact kinetics, and finite size effects \cite{blow2021} which can all be controlled. However, they suffer from computational inefficiency; under conditions where nucleation is very rare, (high temperatures) it is impossible to run brute force simulations for sufficiently long times to see the onset of solidification. 

Seeded simulations can provide parameters for CNT and benchmark other calculations at high temperatures. To avoid the cost of brute force simulations, a seeded simulation introduces a preformed spherical cluster to the supercooled liquid rather than waiting for the clusters to form. Analyzing the evolution of these clusters during subsequent unbiased molecular dynamics simulation provides CNT parameters (see section \ref{sec:seededSimulation} for details) which can then be used with Eq. (\ref{nucRate}) to calculate even the slowest nucleation rates. However, the coarse simplifications of CNT are inherent in these calculations, causing them to fail for systems that do not adhere to the classical pathway \cite{Finney2023}. When a system obeys the CNT assumptions, seeded simulations obtain nucleation rates consistent with more accurate methods \cite{seededLJ, separdar2021a, espinosa2016}.

There are also other methods that can estimate nucleation rates with accuracy between that of brute force and seeded molecular dynamics simulation. These methods bias evolution to explore phase space more efficiently, which reduces computational cost while introducing more sources of error relative to brute force simulations. For comprehensive reviews of molecular simulation methods and their respective advantages,  limitations, and applications to nucleation, see \cite{Sosso2016, blow2021, Finney2023}. In this paper, we will use only brute force and seeded simulations to study solidification of the YOCP due to their simplicity and interpretability.

\section{Brute Force Molecular Dynamics Simulations}\label{sec:bruteForce}
In a brute force simulation a system is initialized in a supercooled liquid state and run with unbiased molecular dynamics until it solidifies. The amount of time required for solidification to occur is determined by the nucleation rate, so we can extract nucleation rates at large supercooling directly from these simulations. 

Note that for our simulations, we used carbon ions at $10^6 \text{g/cm}^3$ (conditions similar to WD). However this is arbitrary because $\kappa$ and $\Theta$ (or $\Gamma$) fully characterize the YOCP. Therefore we present all results in terms of dimensionless quantities that apply to YOCP systems composed of any particles at arbitrary density.

\subsection{ Brute Force Molecular Dynamics Setup}
We perform the molecular dynamics simulations in the Large-scale Atomic/Molecular Massively Parallel Simulator (LAMMPS) \cite{LAMMPS}. Each run has $10^4$ particles of mass $M$ initialized with random positions and initial velocities drawn from a Maxwellian distribution at the appropriate temperature. Since $a$ is a length scale typical of interparticle spacing and the thermal velocity which we define here as $v_{th} = \sqrt{\frac{2k_BT}{M}}$ is a typical particle velocity, $\frac{a}{v_{th}}$ is a time scale for interactions with neighboring particles. We run dynamics with a timestep of $\frac{a}{200v_{th}}$, chosen to be sufficiently small that the total energy of the system is not influenced by numerical error during the integration of the equations of motion. During the MD run at constant volume, a Nose-Hoover thermostat regulates the temperature on a timescale of $t_\text{damp}=\frac{a}{2v_{th}}$ (meaning that if perturbed, the thermostat restores the system to the target temperature in about 100 timesteps) to avoid nonphysical temperature drift. This is important during crystallization because temperature will otherwise increase due to the release of latent heat \cite{papanikolaou2019}. This introduces the assumption that conduction of heat away from a forming crystal is fast relative to the rate at which the crystal forms. This is a limitation, but it is necessary to enable our simulations to run at a fixed thermodynamic conditions.

For the $\kappa=0$ simulations, LAMMPS uses the particle-particle particle-mesh algorithm to carefully handle both the short and long range parts of the coulomb interaction with a relative force accuracy of $\frac{\Delta F}{F}=10^{-5}$ \cite{LAMMPSpppmError}. We found that requiring stricter accuracy tolerances did not affect the time to solidification. The $\kappa = 2$ and $5$ simulations use a Yukawa potential with a cutoff distance of $13.5a$, chosen such that interatomic interactions at distances beyond the cutoff are negligible. 

\begin{figure}
    \centering
    \includegraphics[width=1\linewidth]{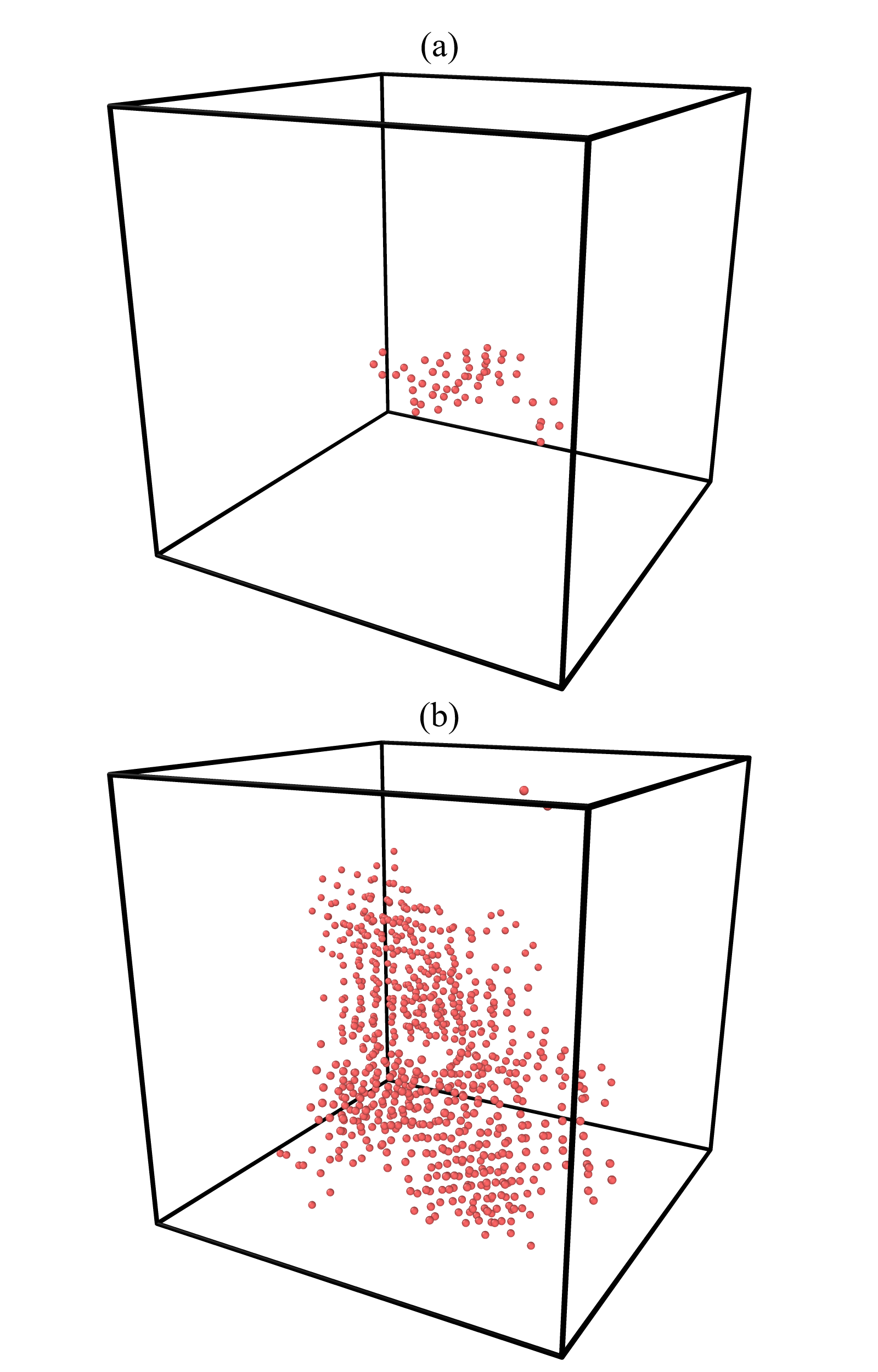}
    \caption{Visualization of the largest cluster in 2 snapshots of a $\kappa=2$, $\Theta = 0.657$ ($\Gamma=670$) simulation at times (a) $t=375 a/v_{th}$ and (b) $t=425a/v_{th}$. Each red dot is a solid particle as identified with polyhedral template matching in OVITO \protect{\cite{Larsen2016, OVITO}}. The clusters around the critical size are compact and nucleation appears to occur in a single step, which is consistent with CNT. They are not spherical when they are small and growing quickly, but become more spherical as they get larger. At higher temperatures (less undercooling) as nucleation gets slower and critical clusters become larger, the assumptions of CNT are more closely obeyed.}
    \label{fig:cluster_vis}
\end{figure}

The simulations were performed at several temperatures for each value of $\kappa$, chosen such that the metastable liquid would remain without large solid clusters for at least $10^4$ time steps, but freezing would begin within $2\times 10^8$ time steps. In practice, this was only possible in a narrow temperature range of $0.62<\Theta<0.73$. (Fig. \ref{fig:phase_diagram}) We ran 3 independent initial configurations for each $(\kappa, T)$ condition. Snapshots during the solidification of one such run are shown in Fig. \ref{fig:cluster_vis}. 

\subsection{Mean First Passage Time Analysis}
\begin{figure}
    \centering
    \includegraphics[width=\linewidth]{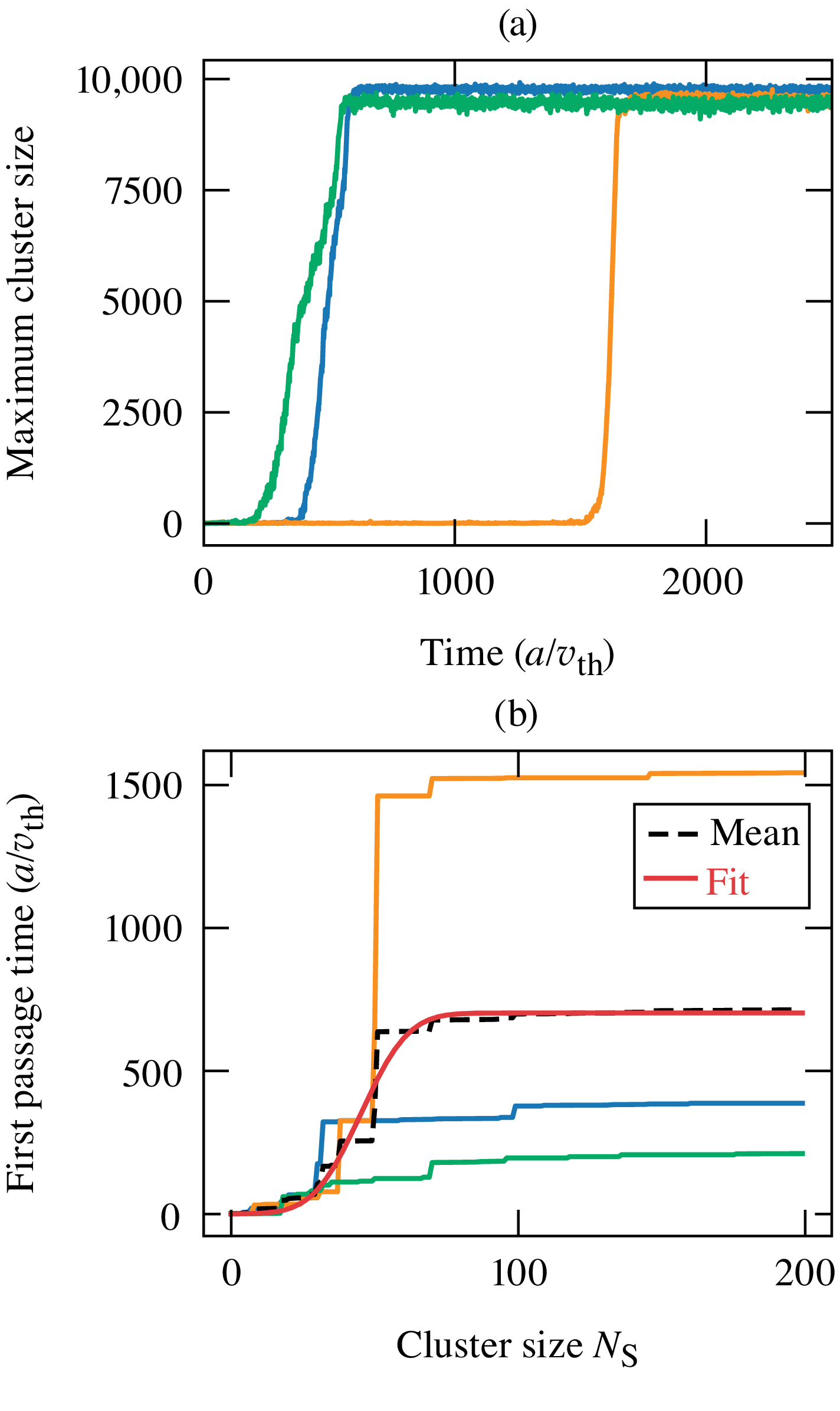}
    \caption{Data from brute force simulations for $\kappa=2$, $\Theta = 0.657$ $(\Gamma=670)$. a) The evolution of maximum cluster size over time for each of the three independent simulations. These show a waiting time before a nucleation event occurs, then fast growth of the cluster to fill the whole volume of $10^4$ particles. b) First passage times (FPT), the time at which each simulation first reaches a particular cluster size, corresponding to the three simulations in (a). The mean first passage time (dashed black curve) from averaging the three independent FPTs and the fit (Eq. (\protect{\ref{MFPT}})), smooth red curve) give an estimated nucleation rate for these conditions}
    \label{fig:bruteForce}
\end{figure}

To analyze our simulation results, we employ the neighbor-averaged Steinhardt bond order parameters $\overline{q}_4(i)$ and $\overline{q}_6(i)$ introduced by Lechner and Dellago \cite{Lechner2008}. These parameters quantify how much 4th and 6th order spherical harmonics contribute to local structure about a chosen ion $i$. Since every crystal structure has known, fixed values for  $\overline{q}_4(i)$ and $\overline{q}_6(i)$ \cite{volkov2002}, we compare the values for each particle in our simulation to known reference values to identify solid-like and liquid-like particles in the simulations. It is noted that a similar order parameter was recently proposed by Klumov \cite{Klumov2021} which is essentially equivalent to the method we used in its ability to accurately identify BCC-like particles.  This analysis proceeds as follows:
\begin{enumerate}
    \item Perform a Delaunay triangulation of the set of atom positions with the qhull library \cite{QHULL};
    \item Generate a list of the neighbors of each ion (the edges in the Delaunay simplices connect neighbor pairs);
    \item Refine each atom's neighbor list by discounting any neighbors that are more distant than 1.2 times the mean distance between the central atom and all identified neighbors. This corrects for degeneracies in the triangulation that can cause bonds to be overidentified in FCC crystals \cite{Brostow1998};
    \item Calculate $\overline{q}_4(i)$ and $\overline{q}_6(i)$ for each atom  (see Ref. \cite{Lechner2008} for details);
    \item Decompose the bond order parameters for each ion into known reference values (from Volkov et al \cite{volkov2002}.) of BCC and FCC crystals by finding values of $C_{FCC}$ and $C_{BCC}$ that minimize \begin{equation*}
\left | \begin{pmatrix}
\overline{q}_4(i)\\ 
\overline{q}_6(i)
\end{pmatrix} 
- C_{FCC} \begin{pmatrix}
0.191\\ 
0.575
\end{pmatrix} 
- C_{BCC} \begin{pmatrix}
0.036\\ 
0.511
\end{pmatrix}\right |
    \end{equation*} with the constraints $0 < C_{FCC}<1$ and $0 <C_{BCC}<1$. If $C_{FCC}<0.5$ and $C_{BCC}<0.5$, the ion is considered liquid-like. Otherwise it is considered a solid-like ion of the phase that contributed more to the decomposition \cite{volkov2002};
\item Group pairs of neighbors that have the same solid-like structure (BCC or FCC) into a cluster.
\end{enumerate}
This method takes a set of ion positions at a particular MD snapshot and generates the total number of solid-like ions, as well as the phase and number of atoms in every cluster in the simulated volume. The result of this analysis for 3 independent runs at the same conditions is shown in Fig. \ref{fig:bruteForce}(a).
Our $\kappa=0$ results (not shown) compare favorably to Fig. 1 in Daligault's brute force simulation study of nucleation in the OCP \cite{daligault2006}.  

We use the mean first passage time (MFPT) method to extract nucleation rate from cluster size data. We invert $N_s(t)$ (Fig. \ref{fig:bruteForce}a) to get the first passage time $\tau(N_s)$, the time at which a cluster of size $N_s$ first appears. Averaging these FPTs for all independent simulations with the same $(\kappa, \Theta)$ gives the MFPT (dashed curve in Fig.\ref{fig:bruteForce}b), which we fit to the expression of Wedekind at al. (see Eq. (8) of \cite{wedekind2007})  \begin{equation}
    \label{MFPT}
    \tau (N_s)=\frac{\tau_{\scriptscriptstyle J}}{2}\left(1+\erf\left\{c\left(N_s-N^*\right)\right\}\right),
\end{equation} where $\erf$ is the error function, for cluster sizes $N_s<200$. The parameter $\tau_{\scriptscriptstyle J}$ is related to the nucleation rate by $J=1/(\tau_{\scriptscriptstyle J}N)$. The fit also provides parameters $c$, which is related to the curvature near the top of the free energy barrier, and $N^*$, the critical cluster size. This is valid as long as nucleation is ``rare'' in the sense that the time required to form a critical cluster ($N^*<200$ for these conditions) is much shorter than the waiting time for crystallization to occur \cite{Nicholson2016}. This is clearly the case for our simulations, as the jump to $N_s \approx 200$ in Fig. \ref{fig:bruteForce}(a) takes place over a short time relative duration for which $N_s\approx 0$ . We confirm that this method gives reasonable results by benchmarking against a previous study in Appendix \ref{app:LJ}.

\section{Seeded Simulations}\label{sec:seededSimulation}
Since brute force MD was only computationally feasible for $\Theta<0.73$, we use seeded simulations at higher temperatures. The simulations are initialized with pre-formed clusters to avoid the long wait times that would be required for them to form spontaneously. Then we extract the CNT parameters $N^*$, $\gamma$, and $D_+$ to estimate nucleation rates and cluster size distributions with less computational cost than with brute force simulations. 

Note that if the bulk chemical potential difference $\Delta \mu$ is known a priori, then running MD simulations to directly determine $N^*$ gives an indirect measurement of the surface contribution to free energy $\gamma$ through  Eq. (\ref{critSize}) \footnote{excluding possible surface curvature dependence}.  Thus $D_+$ and $N^*$ are the only values that need to be extracted directly from seeded simulations.

\subsection{Seeded Molecular Dynamics Setup} \label{sec:seededMD}

We performed MD simulations with seed crystals for $\kappa = 0,2$ and $5$ at reduced temperatures of $\Theta = 0.75, 0.8,$ and $0.85$. Each seeded simulation contained between $N=3\times 10^3$ and $1.6\times 10^4$ particles. We found that results are insensitive to simulation size when the number of particles in the simulation is between $10N^*$ and $30N^*$ (see appendix \ref{app:finite_size} for details), which is consistent with previous studies \cite{seededLJ}. The pair force calculation and the thermostat are the same as for brute force simulations (section \ref{sec:bruteForce}). Seeded and brute force simulations differ in how the positions and velocities are initialized. For each seeded simulation at reduced temperature $\Theta$, we initialize a spherical solid cluster with a particular radius $R$ in a background of supercooled liquid and then allow it to evolve freely with the following procedure \cite{separdar2021a, ruiz2017}:
\begin{enumerate}
    \item Initialize all particles in a BCC lattice ($\kappa = 0,2,5$) or an FCC lattice ($\kappa=5$ only);
    \item  Give all particles random velocities drawn from a Maxwellian distribution at the appropriate temperature;
    \item Run MD on all particles at temperature $T$ for $2\times10^4$ time steps to equilibrate the lattice;
    \item Define a sphere of radius $R$ in the center of the simulation volume. Particles inside the sphere are ``solid'' particles and those outside are ``liquid'' particles;
    \item Over $2\times10^4$ additional time steps, run MD on \underline{only} the liquid particles, ramping their temperature from $T$ to $3T_m$;
    \item  Run MD on the liquid particles at $3T_m$ for $2\times10^4$ time steps;
    \item Over $2\times10^4$ time steps, run MD on the liquid particles, quenching from $3T_m$ to $T_m$;
    \item Over $10^3$ time steps, quench the liquid from $T_m$ to $T$;
    \item Run MD on all particles at $T$ for $5\times 10^4$ time steps. This is the freely evolving part of the run where we output atom positions every 100 time steps;
    \item Trim off all snapshots from the freely evolving run before the temperature of the cluster and liquid have equilibrated.
\end{enumerate}
We use this long equilibration procedure to avoid unwanted stress in the cluster which could affect the final result \cite{zhou2023}.

For the $\kappa = 5$ case, we run MD simulations with both BCC and FCC initial structures. While FCC is the more stable structure, BCC and FCC solids have very nearly the same free energy and it is not clear a priori which phase will nucleate from the liquid first \cite{ostwaldRule}. 

\subsection{Analysis}

We extract the critical cluster size $N^*$ and attachment rate $D_+$ from the seeded simulations. This requires analysis of the growth or decay of the number of solid particles in the cluster over time, $N_s(t)$. This is simple in principle but there are many ways to introduce errors into the calculation. Since the nucleation rate depends exponentially on the size of a critical cluster, these errors can lead to very large uncertainties in the nucleation rate. These large uncertainties were made explicit for the nucleation of NaCl by Zimmerman et al. \cite{Zimmerman2018}.

We improve on previous methods of seeded simulation analysis by choosing a metric of cluster size that comports with assumptions of CNT, accounting for changes to cluster sizes during equilibration, and trying to reduce the influence of statistical noise on the calculation of the attachment rate $D_+$.

\subsubsection{Defining the Cluster Size} \label{sec:cluster_size}

To our knowledge, previous studies that employ seeded simulations \cite{seededAl, seededLJ, ruiz2017, separdar2021a, separdar2021b, espinosa2016, Nederstigt2023, Leysalle2007, bai2005, bai2006} have always identified particles that are part of the cluster using Voronoi polyhedra \cite{Brostow1998, anikeenko2005}, Steinhardt bond order parameters \cite{Lechner2008}, polyhedral template matching \cite{Larsen2016}, or common neighbor analysis \cite{Honeycutt1987}. These methods all rely on measuring some local property to assign a phase to each particle. Since particles near a solid-liquid interface have intermediate values of these local properties, the handling of interface particles is ambiguous and requires arbitrary choices of parameters that affect the number of particles identified in a cluster. This ambiguity leads to uncertainty in cluster sizes of a few hundred particles, but since nucleation rate depends exponentially on the critical cluster size, the uncertainty in the nucleation rate due to cluster identification can be 20 orders of magnitude \cite{separdar2021b,Zimmerman2018}.

Cheng and Ceriotti introduced a collective variable approach to define the number of particles in solid clusters within a liquid \cite{Cheng2015, Cheng2017} which they used for a metadynamics study of diamond nucleation \cite{cheng2022}. Their idea is that in a mixed solid/liquid system, one can compute any extensive order parameter that differs between the bulk solid and bulk liquid phases. The value of this global order parameter for the whole system is the sum of the contributions from the solid and liquid regions. Thus, if expected values for the order parameter in the bulk solid and liquid phases are known, the system can be decomposed into solid and liquid regions without the ambiguity inherent to distinguishing the phases of atoms based on local properties \cite{Cheng2017}.

\begin{figure}
    \centering
    \includegraphics[width=1\linewidth]{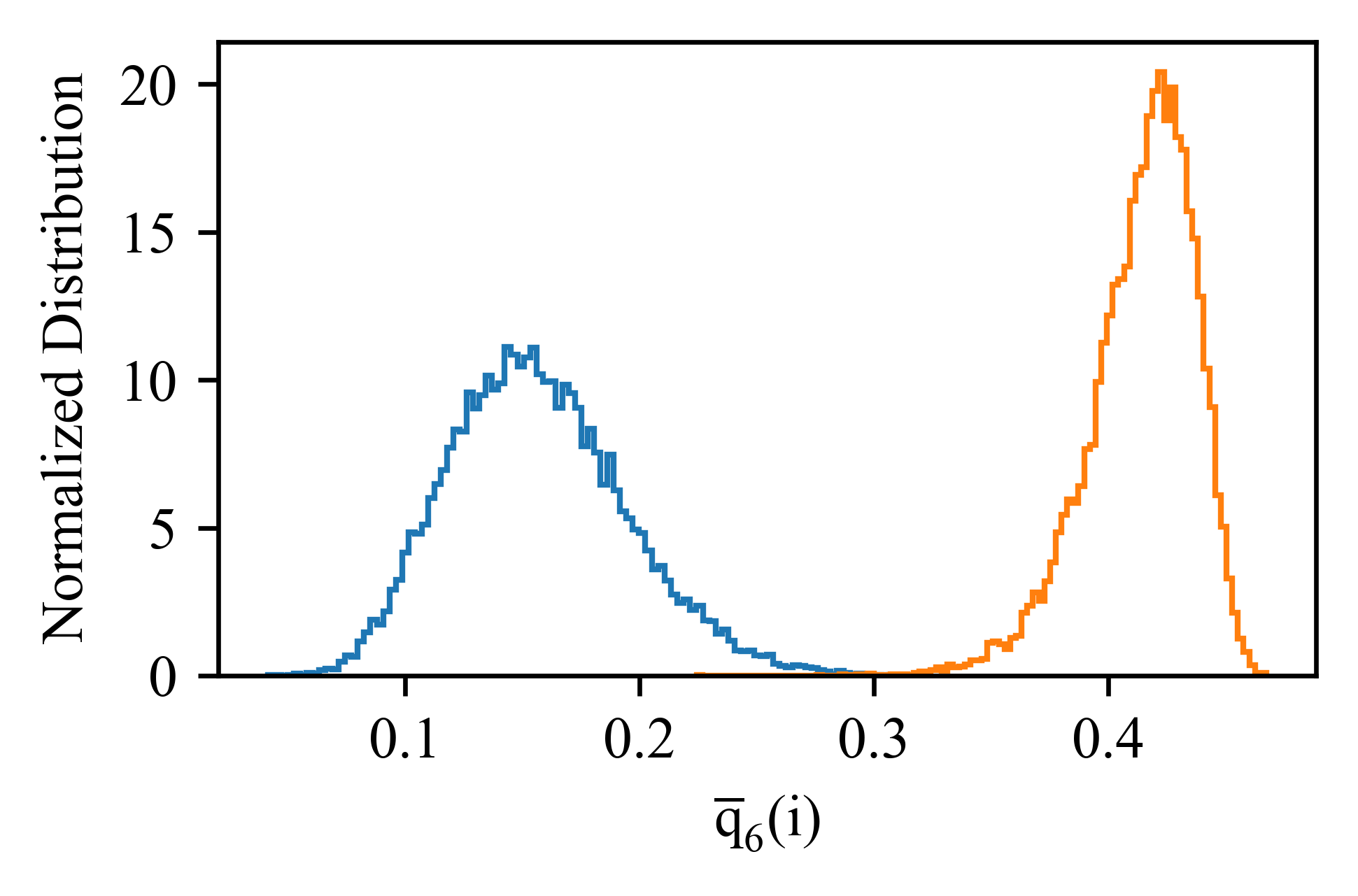}
    \caption{Histogram of the sixth-order, neighbor-averaged Steinhardt bond order parameter, $\overline{q}_6$ in the $\kappa=0$, $\Theta = 0.8$ ($\Gamma = 225$) OCP liquid (blue) and BCC solid (orange). The distributions of $\overline{q}_6$ are distinct between the two phases and vary little over time, which means that the distribution of the collective variable $\overline{Q}_6 = \sum_i \overline{q}_6(i)$ will be very strongly peaked at different values for the solid and liquid and thus provide clear distinction between solid and liquid systems. In a mixed solid-liquid system, the distribution of $\overline{q}_6(i)$ shows two peaks for the solid and liquid components.}
    \label{fig:q6dist}
\end{figure}

We apply this idea by introducing the extensive order parameter $\overline{Q}_6 = \sum_i \overline{q}_6(i)$, summing the previously-used neighbor-averaged Steinhardt bond order parameters over all particles in the simulation. 
$\overline{Q}_6$ is strongly peaked at distinct values in the solid and liquid phases (see Fig. \ref{fig:q6dist}), so we use the simplest decomposition
\begin{equation}
    \label{collectiveQ}
    \overline{Q}_6(t) = \overline{q}_6^\text{liq}N_\ell(t) + \overline{q}_6^\text{sol}N_s(t)
\end{equation}
where $N_\ell(t)$ and $N_s(t)$ are the number of liquid and solid particles in the simulation, respectively, and $\overline{q}_6^\text{liq}$ and $\overline{q}_6^\text{sol}$ are the average values of $\overline{q}_6(i)$ in the bulk phases. This simply means that we assume that of the $N$ total particles, $N_s$ contribute to $\overline{Q}_6$ like the bulk solid, $N_\ell$ contribute like the bulk liquid, and the interface does not contribute at all. These assumptions appear true from our simulations. 

\begin{figure}
    \centering
    \includegraphics[width=1\linewidth]{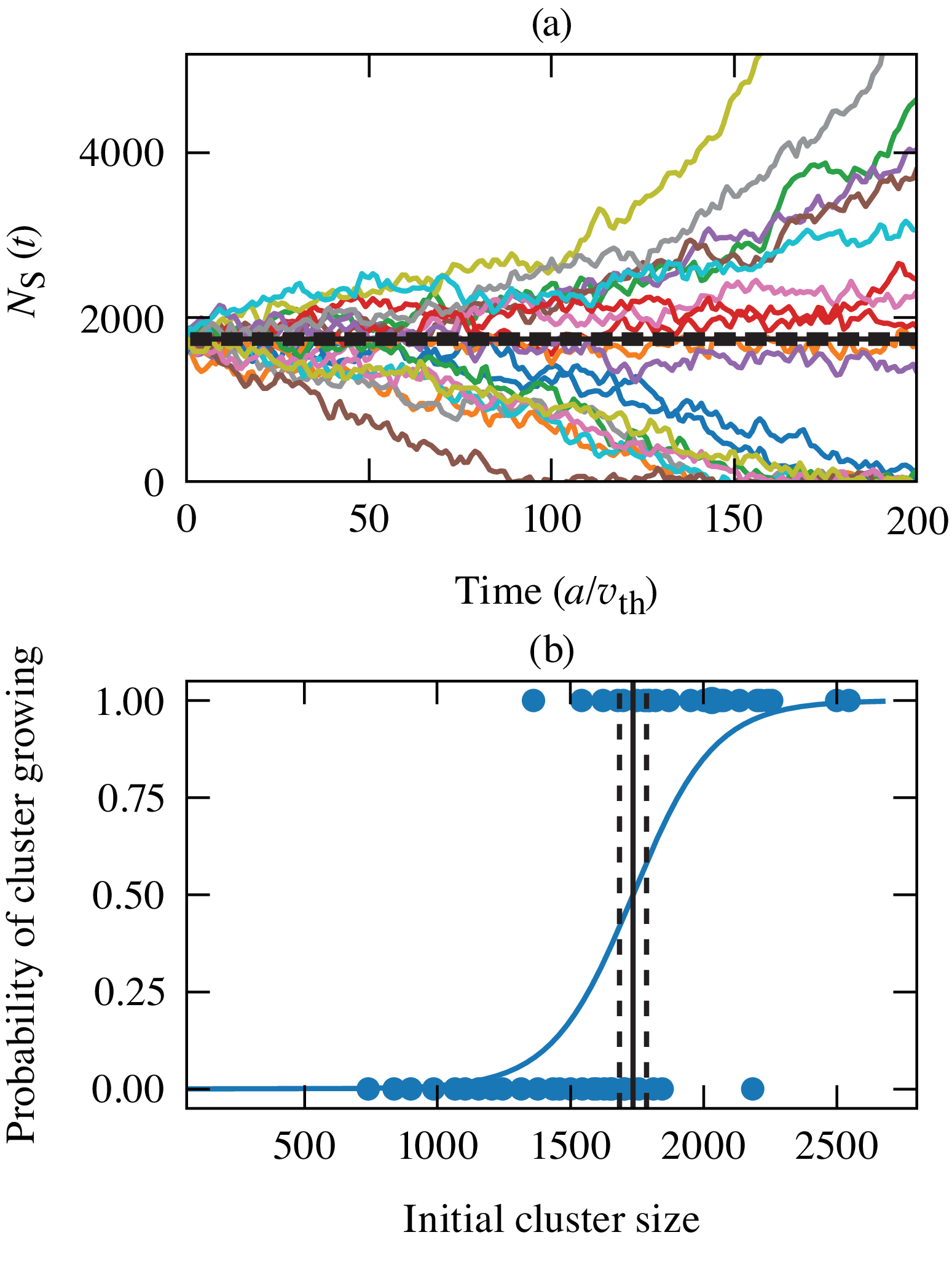}
    \caption{Data corresponding to seeded simulations at $\kappa=2$, $\Theta = 0.85$ $(\Gamma = 518)$. a) The time evolution of cluster size of 20 simulations that start within 10\% of the estimated critical cluster size (black line). Clusters starting above the black line tend to grow and those below tend to shrink, indicating that this is a good estimate of the critical size. b) Probability of a cluster growing as a function of its size. Each point corresponds to one simulation either growing (1) or shrinking (0), so the average value of all points within a range of initial cluster sizes corresponds to a probability of growth. This growth probability is approximated with a fit with Eq. (\protect{\ref{pGrowth}}), yielding an estimated critical cluster size $N^*$ (vertical black line). This is the $N_s$ at which clusters have a 50\% chance of growing and shrinking. The dashed vertical lines represent uncertainties on the estimate of $N^*$ generated from the diagonal elements of the covariance matrix from the fit with Eq. (\protect{\ref{pGrowth}}).}
    \label{fig:seeded_Array}
\end{figure}

For every temperature, we run short simulations of the bulk solid and bulk liquid to calculate the averages  $\overline{q}_6^\text{liq}$ and $\overline{q}_6^\text{sol}$. Then during our seeded simulations we calculate the time-dependent number of particles in the cluster by calculating $\overline{Q}_6(t)$, noting that $N = N_\ell(t) + N_s(t)$, and rearranging Eq. (\ref{collectiveQ}).
\begin{equation}
\label{N_decomposition}
    N_s(t) = \frac{\overline{Q}_6(t) - \overline{q}_6^\text{liq}N}{\overline{q}_6^\text{sol}-\overline{q}_6^\text{liq}}
\end{equation}
See Fig. \ref{fig:seeded_Array}(a) for an example of the $N_s(t)$ that this procedure generates. Fig. \ref{fig:seeded_Array} shows that the larger clusters usually grow and smaller clusters usually shrink, which is consistent with the predictions of CNT.

The major assumption of this method is that the system can be clearly divided into solid-like particles inside the cluster and liquid-like particles outside. However, this is already implicit in our analysis because we use CNT which requires that the cluster has the properties of the bulk solid, the liquid has the properties of the bulk liquid, and the interface between the two is infinitesimally thin. Therefore, this cluster size identification method does not introduce additional error into our calculation beyond reliance on the approximations inherent to CNT. Unlike previous seeded simulation studies, this method of computing cluster sizes does not rely on arbitrary parameters. 

\subsubsection{Determination of the Critical Cluster Size $N^*$}

Calculating the critical cluster size is straightforward in principle. When a cluster is smaller than $N^*$ it tends to shrink and when it is larger it tends to grow. This allows us to estimate $N^*$ by finding the number of particles in a cluster that is equally likely to grow or shrink during subsequent MD simulation. Previous studies typically generate many clusters with the same initial radius and then adjust the temperature until a 50\% probability of growth is achieved \cite{bai2005,separdar2021a, separdar2021b, zhou2023, espinosa2016, seededLJ}. In practice, 5 to 10 initial cluster configurations are run at many different temperatures until a temperature is found that causes, for example, 3 of 6 clusters to grow. This method does not give precise estimates of the critical size because the temperature is usually only tested in increments of about 1\% (and a 1\% temperature difference can change $N^*$ by hundreds of particles), 10 trials is not sufficient to be confident that the growth probability is 50\%, and the ensemble of clusters supposedly initialized at the critical size actually start with a variety of different initial sizes due to drift during equilibration. We remedy all of these issues.

For each temperature, we manually choose by trial-and-error several initial cluster radii above and below the critical radius. 
For each of these radii, we run at least 10 simulations with the procedure outlined in section \ref{sec:seededMD}. Each of the 10 simulations starts with the same cluster size, but the particle velocities are initialized differently which causes cluster sizes to differ after equilibration. Then, each cluster either grows or shrinks between the end of equilibration and the end of the simulation. We assign each simulation a value of 0 if its cluster shrinks or 1 if it grows. The average of these values for all simulations that started within a particular range of cluster sizes approximates the growth probability of clusters within that size range. 

We plot these values for each individual simulation in Fig. \ref{fig:seeded_Array}(b) and fit to 
\begin{equation}
    \label{pGrowth}
    P_\text{growth}(N_s) = \frac{1}{1+e^{C(N_s-N^*)}}
\end{equation}
where $C$ and $N^*$ are parameters of the fit with $N^*$ being the critical cluster size. This functional form has the correct limiting behavior ($P_\text{growth}(N_s>>N*) = 1$, $P_\text{growth}(N_s<<N*) = 0$), and it satisfies $P_\text{growth}(N_s=N^*) = 0.5$ i.e. a cluster size of $N^*$ has an equal probability of growing or shrinking. 

We find that this method is robust and only requires around 50 simulations to densely sample cluster sizes around $N^*$ to achieve a converged estimate.  This method does not have important temperature uncertainty, uses all simulations to make a good estimate of $N^*$, and where cluster size fluctuations during equilibration was a liability for previous studies, this method uses that variation to more densely sample possible $N^*$ values.
\begin{figure}
    \centering
    \includegraphics[width=1\linewidth]{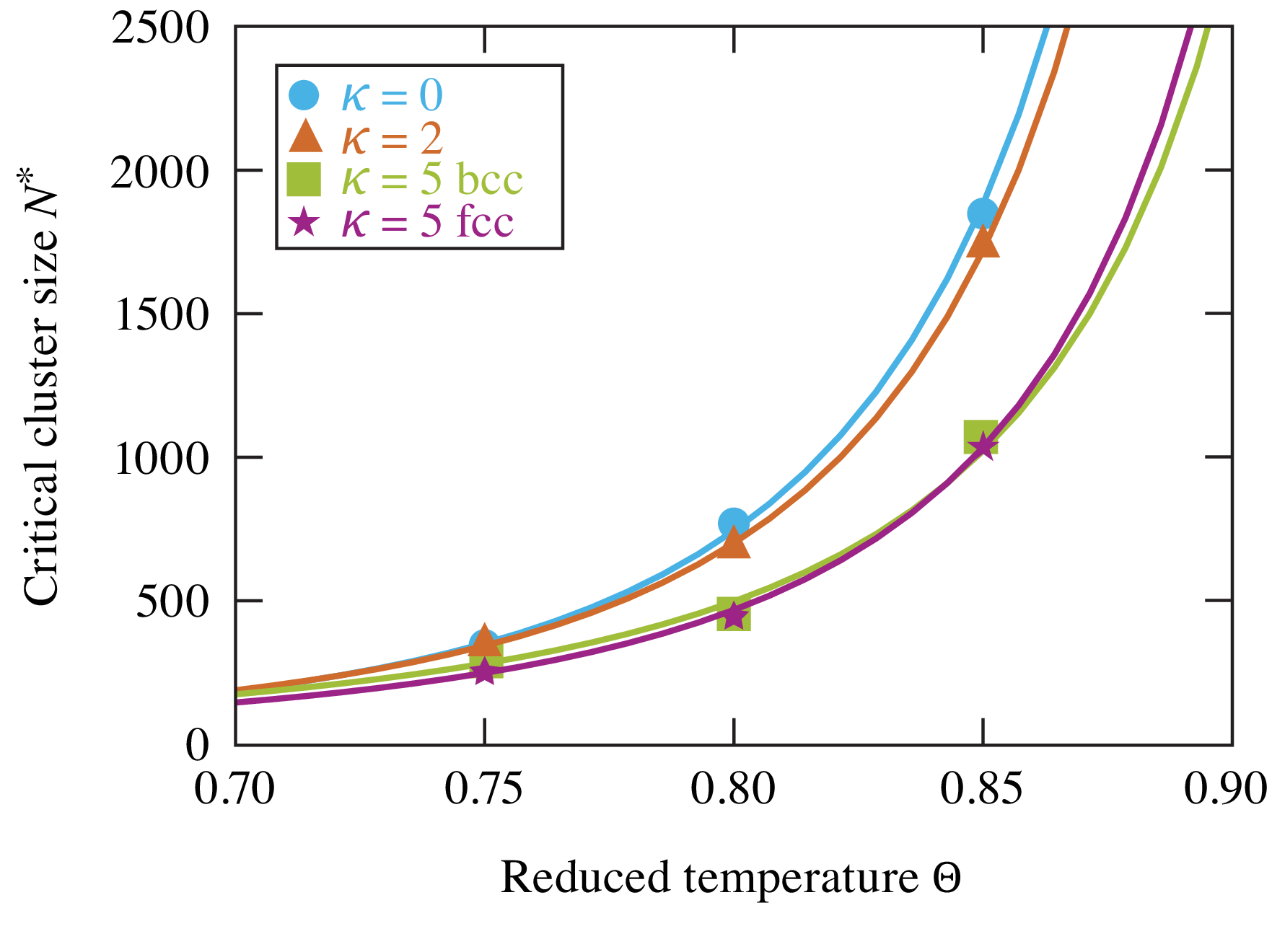}
    \caption{Critical cluster sizes estimated from seeded simulations. Points were generated directly from simulations as shown in Fig. \protect{\ref{fig:seeded_Array}}. Increasing screening $(\kappa)$ decreases $N^*$ at fixed $\Theta$, but using a BCC or FCC initial cluster does not affect the growth in the $\kappa=5$ simulations. The critical cluster size diverges as $\Theta\rightarrow1$ (i.e. $T \rightarrow T_m$)  because there is no driving towards solidification $(\Delta \mu = 0)$ when $\Theta=1$. The curves come from Eq. (\protect{\ref{critSize}}) combined with the linear fits to $\gamma$ shown in Fig. \protect{\ref{fig:interfacialEnergy}}.}
    \label{fig:nCrit}
\end{figure}

The fit generated with this procedure is shown in Fig. \ref{fig:seeded_Array}(b) and the values of $N^*$ calculated with this method are shown in Fig. \ref{fig:nCrit}. Our estimated critical cluster sizes are around 200 to 2000 for the $(\kappa, \Theta)$ conditions we covered with our seeded simulations. These values are reasonable, as other systems at $\Theta = 0.75$ like Lennard-Jones fluids and liquid aluminum at have been observed to have $N^*\approx 250$ and $1000$, respectively \cite{Moroni2005, Mahata2018}.

\subsubsection{Obtaining the Bulk Chemical Potential Difference $\Delta \mu$}
Applying CNT to crystallization of the supercooled YOCP requires the difference between the chemical potentials for the solid and liquid phases for all relevant temperatures and $\kappa$: $\Delta \mu = \Delta f + P \Delta v$ where $\Delta f$ and $\Delta v$ are the specific (per particle)  Helmholtz free energy and volume, respectively.  In dense astrophysical plasmas, but not in general, $\Delta v$ is very small and we use $\Delta \mu \sim \Delta f$. We use free energies from Ref. \cite{Farouki1994} ($\kappa=0$) and Ref. \cite{HamaguchiTriplePoint1997} ($\kappa = 2,5$), which give separate expressions for the liquid and solid \footnote{\cite{HamaguchiTriplePoint1997} does not provide fits to BCC free energy at $\kappa=5$. For this case, we assume that $\Delta \mu_{BCC} = \Delta \mu_{FCC}$, which is reasonable because BCC and FCC energies are very close and we observe a mix of BCC-like and FCC-like particles in the simulations}. Since we are interested in the supercooled liquid, we extrapolate the expressions for the liquid free energy to temperatures below the temperature of the simulations used to generate them. This is reasonable because the functional form of the fit is chosen to work well for the OCP below the melt temperature \cite{stringfellow1990} and the excess energy fits are clearly correct at low temperatures because they converge to the Madelung energy \cite{HamaguchiTriplePoint1997}.

\subsubsection{Calculating the Surface Free Energy Coefficient $\gamma$}

\begin{figure}
    \centering
    \includegraphics[width=1\linewidth]{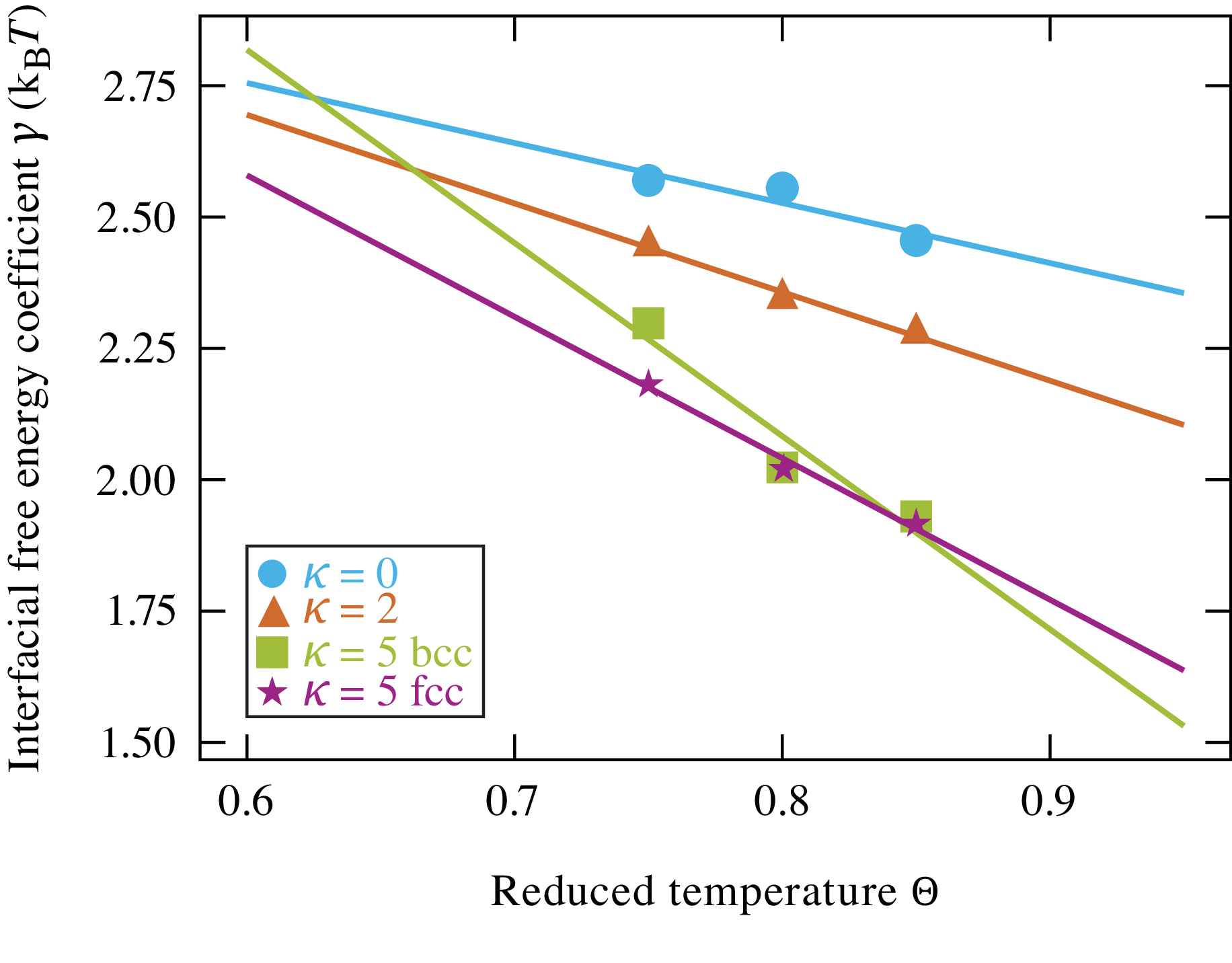}
    \caption{Free energy coefficient $\gamma$ associated with the interface between spherical critical clusters and a liquid background. Increasing $\kappa$ decreases the free energy associated with the interface, which contributes to strongly screened systems having a smaller critical cluster size. The lines are linear least-squares fits.}
    \label{fig:interfacialEnergy}
\end{figure}

With estimates of $N^*$ and $\Delta \mu$, we can calculate $\gamma$, the coefficient of the interfacial contribution to cluster free energy by inverting Eq. (\ref{critSize}) (Fig. \ref{fig:interfacialEnergy}). We find that Yukawa systems with weaker screening (smaller $\kappa$) have larger free energy penalties associated with solid-liquid interfaces.

Since BCC and FCC structures have very similar bulk free energies for $\kappa = 5$, the difference in $\gamma$ between BCC and FCC initial clusters comes from differences in their critical cluster size. Specifically, BCC initial clusters have a larger $N^*$ than FCC clusters at $\Theta = 0.75$, so the BCC clusters must have a correspondingly larger $\gamma$.

\subsubsection{Calculating the Attachment Rate $D_+$} \label{sec:attachment}

\begin{figure}
    \centering
    \includegraphics[width=1\linewidth]{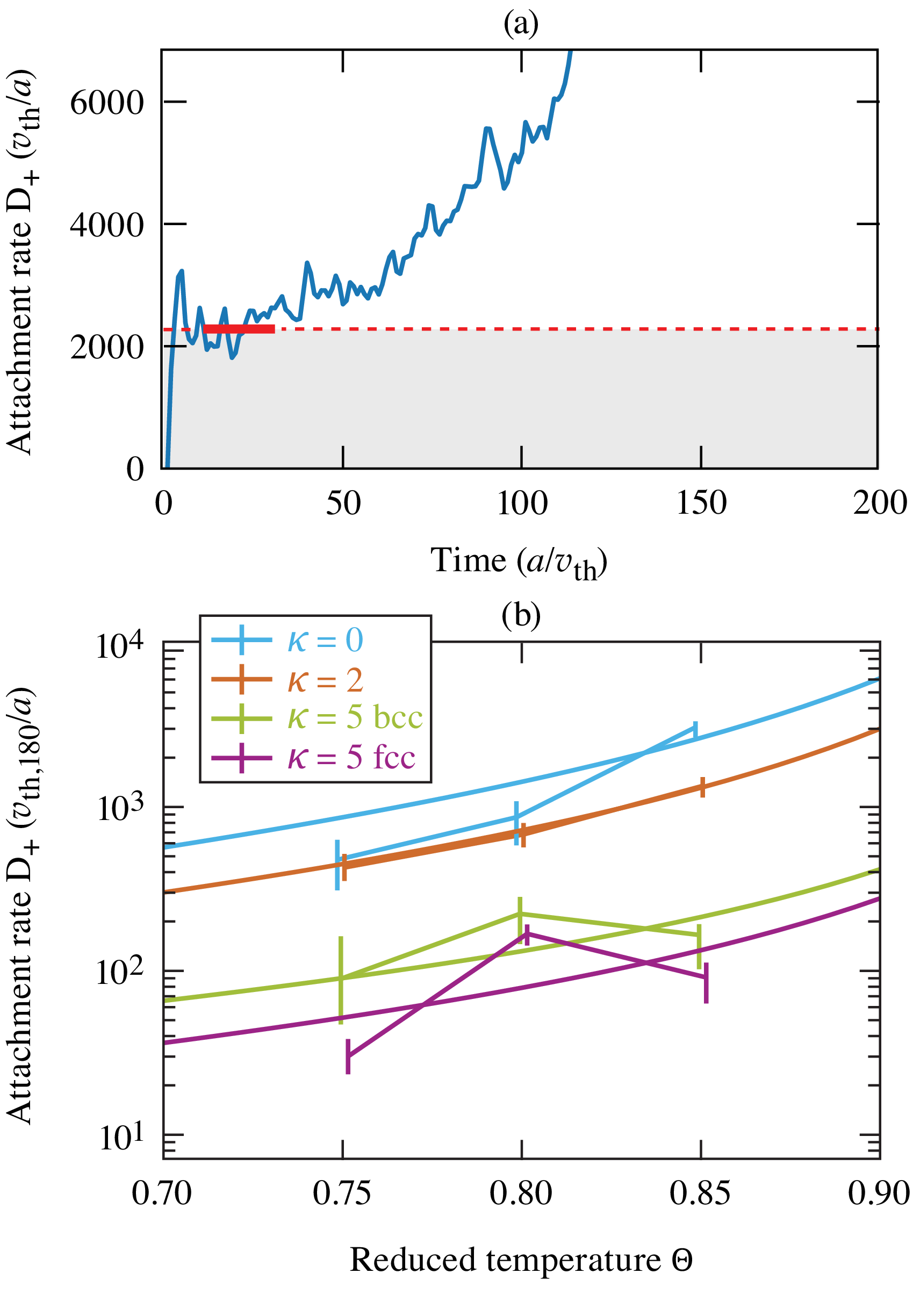}
    \caption{a) Attachment rate $D_+$ calculated from the $\kappa =2$, $\Theta=0.85$ data in Fig. \protect{\ref{fig:seeded_Array}}(a) using Eq. (\protect{\ref{attachment}}). The attachment rate is given by the spread in the $N_s(t)$ data because when a system crosses its nucleation barrier diffusively, the attachment rate acts as a diffusion coefficient governing the evolution of the cluster size \protect{\cite{ruiz1997, Auer2004}}. The initial peak is caused by noise in $N_s(t)$ data. The solid portion of the horizontal red line is the time over which we averaged this curve to obtain an estimated $D_+$. The height of the red line indicates the estimate for this example.  b) Attachment rates calculated from seeded simulations, as shown in (a). For easy comparison, all attachment rates are shown in the same time units determined by the thermal velocity for a plasma at $\Gamma =180$, $v_{th.180}$. The smooth curves are least-squares fits with Eq. (\protect{\ref{eq:attachmentFit}}).}
    \label{fig:attachmentRate}
\end{figure}

The time evolution of each seed cluster, summarized by $N_s(t)$, contains the information we need to calculate the attachment rate $D_+$. We assume that attachment and detachment of ions from the cluster occurs frequently and randomly such that cluster sizes near the critical size evolve through a random walk. This idealization, called diffusive barrier crossing, has been observed to be a reasonable description of nucleation in MD simulations \cite{Moroni2005}. In this regime, fluctuations in cluster sizes are determined by the rate of attachment and detachment events \cite{ruiz1997}. In particular, the attachment rate is like a diffusion coefficient that characterizes how quickly an ensemble of clusters starting near $N^*$ spreads to different cluster sizes. Thus we can calculate $D_+$ with an average over seeded simulations \cite{Auer2004}:
\begin{equation}
    \label{attachment}
    D_+ = \frac{\left\langle \big( N_s(t) - N_s(0) \big)^2 \right\rangle}{t}.
\end{equation}
We perform this average over all simulations for which $N_s(0)$ is within 10\% of $N^*$. The result is a series of estimated values of $D_+$ as a function of $t$. This is shown in Fig. \ref{fig:attachmentRate}(a). There is often a spike at short times due to noise in $N_s(t)$ which must be excluded from the estimate of $D_+$. See Appendix \ref{app:finite_size} for discussion of the influence of noise. We also must exclude large times when $D_+$ diverges due to clusters catastrophically growing or shrinking when they get far from $N^*$. To account for this, we average $D_+$ over times from 10 to 30 $a/v_{th}$ because we observed that this time range falls after the initial spike and before the later divergence for all conditions. 

These final values are plotted in Fig. \ref{fig:attachmentRate}(b). These results show that attachment rates generally increase with temperature for all $\kappa$. This happens because $N^*$ increases with temperature so the critical cluster has more surface area to which the ions can attempt to attach and because increased thermal motion allows more attempted attachments per unit surface area. The attachment rate is small for larger $\kappa$ mostly because the melting point of the strongly screened YOCP is much colder than the OCP \cite{HamaguchiTriplePoint1997} so mobility of the strongly screened ions is smaller at fixed $\Theta$.

To extrapolate these attachment rates to temperatures beyond those those of our seeded simulations, we use a physically motivated functional form. A common approximation for the attachment rate is $D_+ \propto \frac{D {N^*}^{2/3}}{\lambda^2}$ where $\lambda$ is the distance a particle must move to attempt to attach to the cluster and $D$ is the bulk diffusion coefficient \cite{Kelton1991}. Previous work indicates that for Yukawa plasmas with screening parameters up to $\kappa=5$, the temperature dependence of the bulk diffusion coefficient is approximately given by $D(T) \propto T^2$ for strong coupling $\left(\Theta<10\right)$ \cite{khrapak2018}. Assuming $\lambda$ is independent of temperature gives the functional form for the attachment rate 
\begin{equation}
    \label{eq:attachmentFit}
    D_+ = D_0 \Theta^2 {N^*}^{2/3},
\end{equation} where $D_0$ is the only parameter. The resulting least-squares fits are shown in Fig. \ref{fig:attachmentRate}(b), and our values for $D_0$ are in Table \ref{tab:fit_params} of appendix \ref{app:fits}.

\section{Results/Discussion} \label{sec:results}
\subsection{Qualitative Aspects of Nucleation}

We observe that $\kappa = 0$ and $2$ simulations always result in BCC crystals while $\kappa = 5$ forms either BCC or FCC structures, with FCC appearing to be more common at lower temperatures. This is in general agreement with the YOCP phase diagram, although FCC is the equilibrium solid phase for $\kappa=5$ \cite{HamaguchiTriplePoint1997}. It is interesting but not unprecedented that we observe fast BCC formation in systems where FCC is the most stable phase. General theoretical considerations support a BCC precursor phase \cite{klein1986, Alexander1978} and MD simulations of rapidly quenched Lennard-Jones \cite{tenWolde1996} and aluminum \cite{papanikolaou2019} fluids have shown BCC formation alongside a different thermodynamically-favored bulk phase

In all simulations, the initial supercooled liquid is unstructured with broad peaks in the radial distribution function (RDF) that decay quickly with distance, indicating little long-range order. The final configuration is highly ordered; the RDF develops peaks at larger distances and the peaks have substructure characteristic of the BCC (or FCC) lattice.

\begin{figure}
    \centering
    \includegraphics[width=1\linewidth]{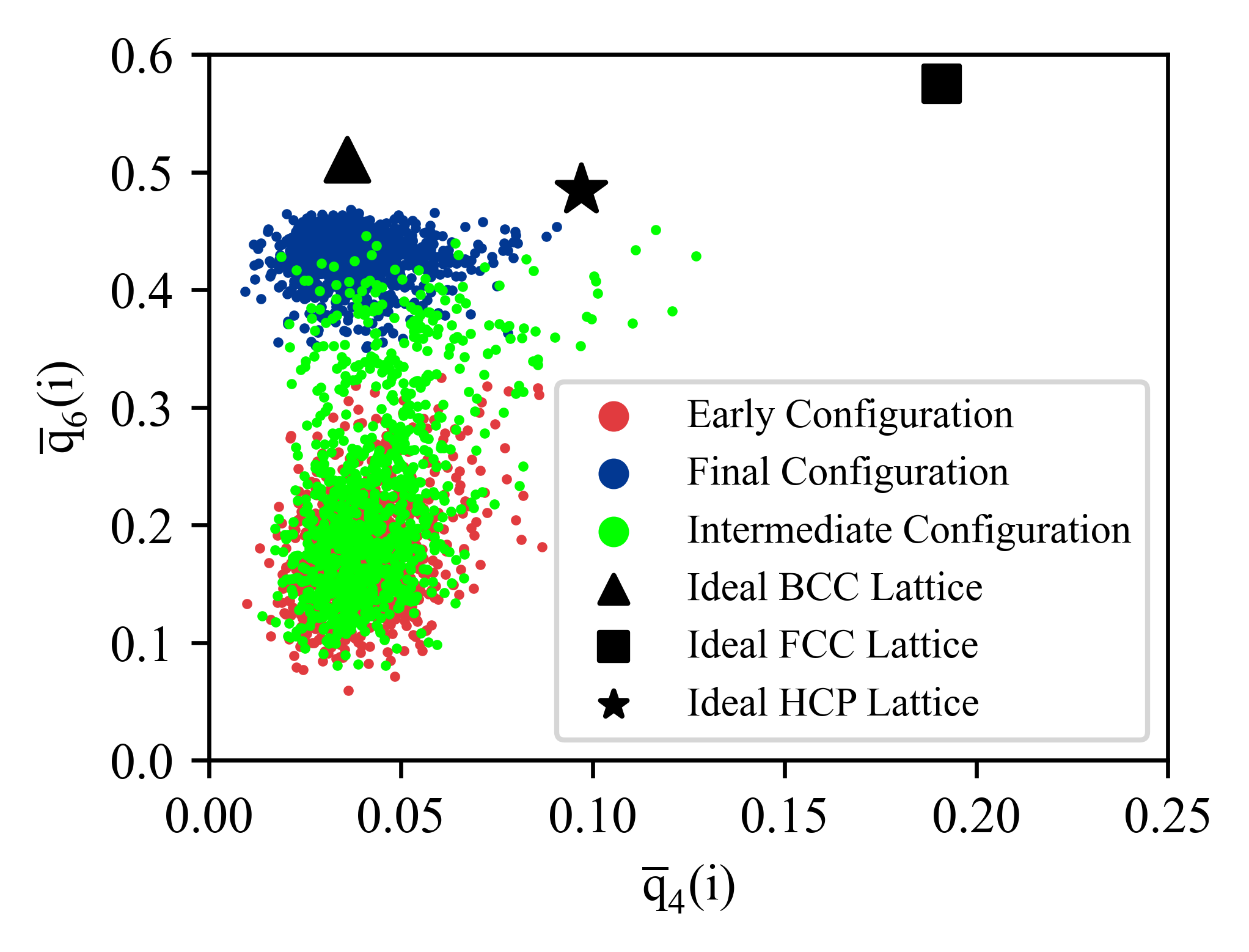}
    \caption{Distributions of particles in the $\overline{q}_4\text{--}\overline{q}_6$ plane during a brute force simulation, calculated with the Voronoi defined neighbors in section \protect{\ref{sec:bruteForce}}. Red and blue points show the distributions before the initiation of solidification and after the whole simulation volume is solidified, respectively. These distributions match those obtained by equilibrating the bulk solid or liquid. The intermediate distribution (green points) is generated from the configuration shown in the bottom panel of figure \protect{\ref{fig:cluster_vis}}. This configuration has solid-like and liquid-like particles which correspond to the green points overlapping the red and blue distributions. For clarity, only $1/10$ of the total particles are shown in this plot.}
    \label{fig:BOP_evolution}
\end{figure}

Time evolution of the neighbor-averaged bond order parameters follows the same qualitative trends discussed above. Figure \ref{fig:BOP_evolution} displays the distribution of particles in the $\overline{q}_4\text{--}\overline{q}_6$ plane for three snapshots during a brute force simulation. At early times the distribution matches an equilibrated liquid and at late times the distribution matches a BCC solid. At the intermediate snapshot, a crystallite has formed but not filled the full simulation volume. This is reflected in the  $\overline{q}_4\text{--}\overline{q}_6$ distribution, which overlaps with both the bulk solid and bulk liquid distributions. Interestingly, some particles reach slightly higher values of $\overline{q}_4$ during solidification than are reached in either the solid or liquid states.

\subsection{General Results}

\begin{figure}
    \centering
    \includegraphics[width=1\linewidth]{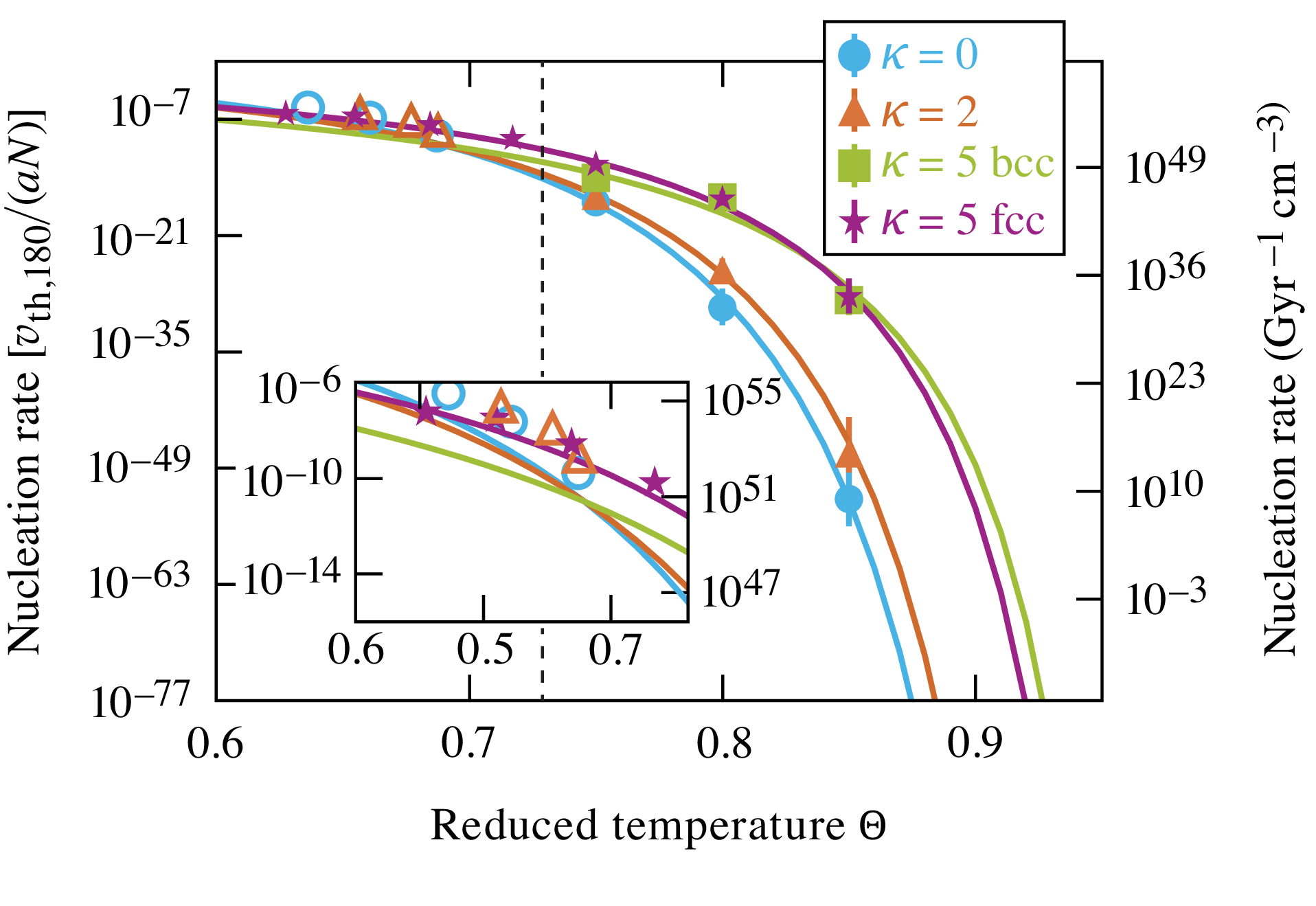}
    \caption{Homogeneous nucleation rates derived from brute force and seeded simulations. Filled points to the right of the dashed line come from Eq. (\protect{\ref{nucRate}}) with parameters determined from seeded simulations. The curves extrapolate the seeded simulations using the fits shown in Figs. \protect{\ref{fig:interfacialEnergy}} and \protect{\ref{fig:attachmentRate}}(b). These extrapolations agree with brute force simulations, (open points to the left of the dashed line) showing that CNT describes the temperature dependence of the nucleation rate in the range $0.62 \le \Theta \le 0.85$. This agreement is also showed in more detail in the inset plot. The left y-axis is general for any Yukawa plasma. The right y-axis provides physical context in the specific case of WD applications. Assuming a $10^6 \text{g/cm}^3$ carbon Yukawa plasma, the right axis gives the nucleation rate in nucleation events per $\text{cm}^3$ per Gyr. When $\Theta \gtrsim 0.9$, it would take many billions of years for a nucleation event to occur within a cubic centimeter and therefore classical homogeneous nucleation may not be relevant for initiating WD solidification.}
    \label{fig:nucRate}
\end{figure}

With the values of $N^*$, $\gamma$, and $D_+$ extracted from our seeded simulations, along with Hamaguchi's values for $\Delta \mu$, Eq. (\ref{nucRate}) gives the nucleation rates for $\Theta \ge 0.75$. Fig. \ref{fig:nucRate} shows the nucleation rates from the simulations and extrapolations generated from the fits to $N^*$ (Fig. \ref{fig:nCrit}), $\gamma$ (Fig. \ref{fig:interfacialEnergy}), and $D_+$ (Fig. \ref{fig:attachmentRate}). See Appendix \ref{app:fits} for the fit parameters that produce these curves. At low temperatures ($0.62 \le \Theta \le 0.73$), the extrapolated analytic form agrees well with the brute force data. This agreement indicates that single-step, classical nucleation is a reasonable description of the nucleation pathway in the supercooled YOCP and that even at low temperatures, the assumptions of CNT are justified for a quantitative description of nucleation. Note that this fit will not be valid for temperatures much lower than our brute force simulations because the YOCP is expected to form a glass around $\Theta = 0.5$ \cite{Ogata1992,Castello2021, Yazdi2014}. 
 
For all $\kappa$, increasing temperature closer to $\Theta=1$ leads to much slower nucleation. This occurs because the critical cluster size $N^*$ increases with temperature and the nucleation rate is exponentially suppressed with increasing $N^*$. For example, when the $\kappa=0$ OCP increases in temperature from $\Theta=0.8$ to $\Theta=0.81$, our estimated $N^*$ increases from $744$ to $880$, which leads to a decrease in the nucleation rate of more than 3 orders of magnitude. 

Figure \ref{fig:nucRate} also shows that as screening becomes stronger the nucleation rate increases at fixed $\Theta \gtrsim 0.7$. This happens because systems with weak screening have longer range interactions and similarly longer correlation lengths. This long correlation length increases interfacial free energy because particles near the interface ``see'' many particles of a different, incongruous phase. The high interfacial free energy leads to a larger critical cluster size which require larger (less probable) departure from the bulk liquid phase to form. Therefore nucleation can occur more frequently in systems with strong screening. Note, however, that this is only true at fixed reduced temperature $\Theta$. At fixed absolute temperature $T$, the $\kappa=5$ YOCP is less supercooled than the $\kappa = 0,2$ cases (due to the $\kappa$ dependence of $T_m$) and will therefore always have a lower nucleation rate \cite{HamaguchiTriplePoint1997}.

\begin{figure}
    \centering
    \includegraphics[width=1\linewidth]{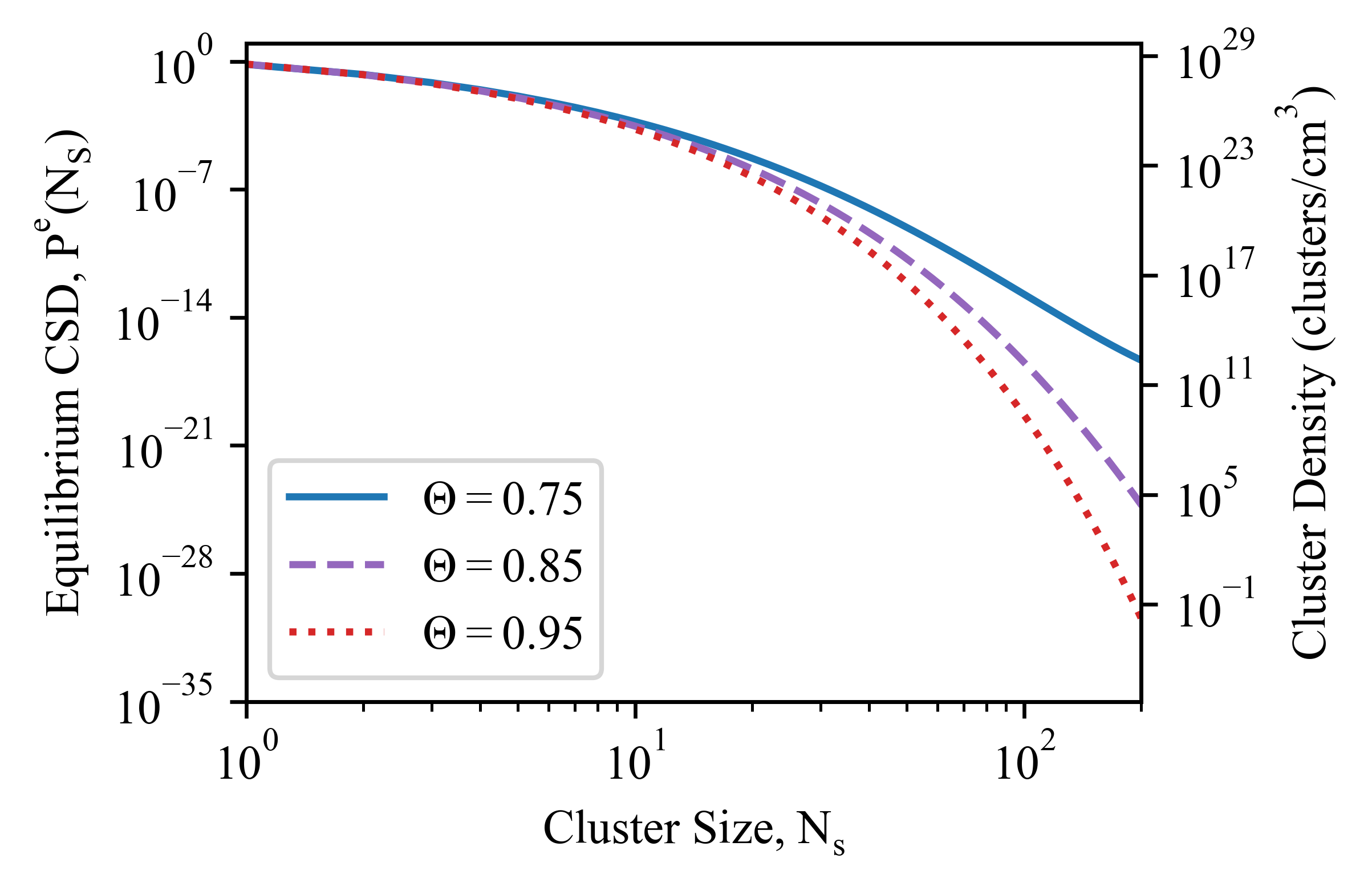}
    
    \caption{Equilibrium Cluster size distribution (CSD) estimated from seeded simulations and an assumed quasi-Boltzmann distribution [Eq. (\protect{\ref{distEq}})] showing only cluster sizes for which $\Delta G(N_s)<<\Delta G(N^*)$ where the equilibrium expression is valid. $P^e(N_s)$ should be understood as fraction of clusters in a liquid system with size $N_s$. The different curves show that as temperature increases, the prevalence of large clusters decrease. The right y-axis provides physical context for a $10^6 \text{g/cm}^3$ pure carbon plasma. It shows the number of clusters of each size expected to be present in a cubic centimeter. This shows that even around the melting temperature when homogeneous classical nucleation is too slow to cause solidification, transient clusters of $\sim 100$ particles are prevalent, but represent a very small fraction of the total number of particles.}
    \label{fig:clusterDist}
\end{figure}

We now evaluate the CNT equilibrium cluster size distribution [Eq. (\ref{distEq})] which is valid for small clusters deep in the $N<N^*$ free energy well. Fig. \ref{fig:clusterDist} shows these distributions for $\kappa=0$ (close to WD conditions) for several temperatures. These cluster size distributions, $P^e(N_s)$, represent the probability that a cluster is of size $N_s$. Previous studies have shown that this CNT $P^e(N_s)$ approximates the actual distribution of small nuclei in large brute force MD simulations \cite{Swope1990, Auer2004, Porion2024}.

These cluster size distribution show that small clusters occur frequently, but large clusters rarely occur due to their large free energy of formation. They also show that as temperature decreases, clusters of all sizes become more common due to the solid becoming increasingly thermodynamically favorable.

As $\kappa$ becomes larger with fixed $\Theta$, all cluster sizes become more common. This means that large clusters should become more common as the YOCP approaches the strong screening limit due to the smaller free energy penalty associated with the interface.  Note that, exactly like the nucleation rate, this applies only at fixed \emph{reduced} temperature. At a fixed absolute temperature, the strongly screened YOCP is less undercooled than the weakly screened YOCP, which makes cluster formation less favorable with strong screening. 

These results all depend on the reliability of CNT. Agreement between seeded and brute force simulations indicate that CNT may be reasonable, but there are still many ways that CNT could fail at temperatures near the melt temperature. 

\subsection{Application to White Dwarf Stars } \label{sec:WD_results}

Stars with an initial mass below 8 solar masses will end their lives as white dwarfs. Without the energy generated by nuclear fusion, white dwarfs evolve by simply cooling for billions of years. The crystallization of the strongly coupled plasma at the core is an important phase in the late cooling of white dwarfs. Models of the cooling of white dwarf stars assume that the dense plasma freezes instantaneously once it reaches a temperature below $T_m$. This is justified intuitively by the very long cooling time scale of the order of a billion years (Gyr). On the other hand, our calculations of the classical homogeneous nucleation rate indicate that the plasma must cool further before freezing.  To demonstrate the potential relevance of our results to white dwarf stars, we consider the delay in the onset of crystallization of the core due to requiring various degrees of undercooling.

For this purpose, we consider the cooling of a white dwarf model of total mass $M_\star=0.6$ solar mass, with a pure carbon core overlain by a $10^{-2}\,M_\star$ layer of helium and a $10^{-4}\,M_\star$ layer of hydrogen. These parameters are representative of the most common white dwarf stars\footnote{Although actual white dwarf stars have cores composed of comparable masses of carbon and oxygen, we choose pure carbon here for consistency with our YOCP simulations.}. The crystallization delay for a given degree of undercooling can be estimated by comparing the standard cooling sequence that starts to crystallize at the equilibrium value of $T_m$ to a calculation where crystallization is artificially inhibited and the core remains liquid until it reaches some lower temperature \cite{STELLUM}. The crystallization onset delay is the time that elapses between when the standard model begins to crystallize and when the model without crystallization reaches a temperature with a particular nucleation rate. 

\begin{figure}
    \centering
    \includegraphics[width=1\linewidth]{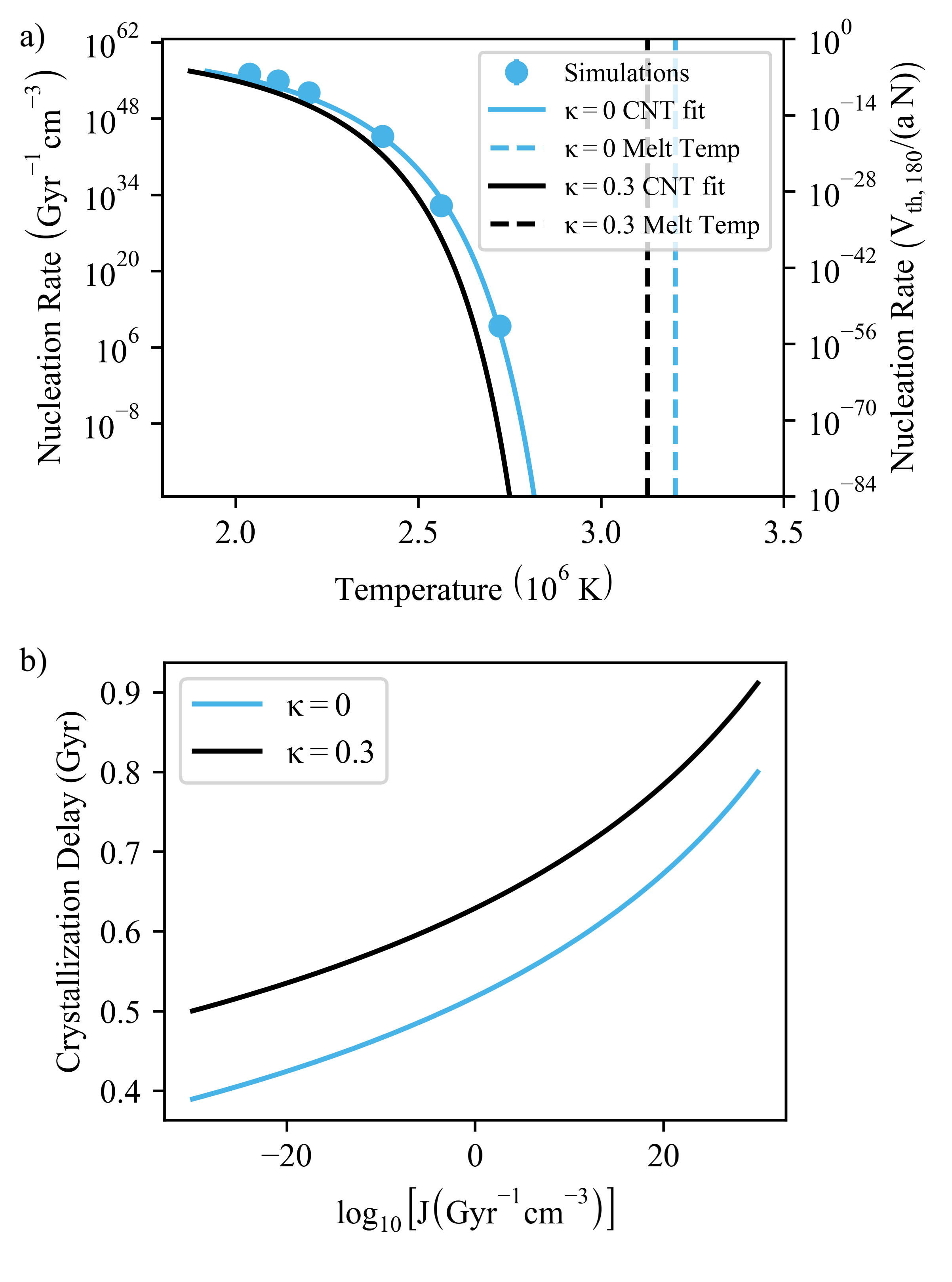}
    \caption{a) The nucleation rate of a supercooled $3.65 \times 10^6 \text{g}/\text{cm}^3$ pure carbon YOCP for $\kappa=0$ (blue) and $\kappa=0.3$ (black). These nucleation rates are too small to cause solidification of the star's core within $\sim$10\% of $T_m$ (dashed lines) meaning that there is either a delay in crystallization until $\sim$10\% supercooling is reached, or there is a nonclassical nucleation mechanism accelerating WD crystallization. b) Crystallization delays, the time required for the star to cool from $T_m$ to the temperature at which the nucleation rate is sufficiently to initiate crystallization.}
    \label{fig:WD_cooling}
\end{figure}

We recast the nucleation rates of Fig. \ref{fig:nucRate} in more appropriate units in Fig. \ref{fig:WD_cooling}a for $\kappa=0$ and 0.3 \footnote{ The latter is obtained by a temperature rescaling of the $\kappa=0$ curve to capture the screening from the relativistic degenerate electron background.}. These curves show that while $T_m = 3.09 \times 10^6 \text{K}$, there is fewer than 1 nucleation event per billion years in the entire WD until the star cools much further, to around $2.7 \times 10^6 \text{K}$. This means that we do not expect crystallization until the WD core is undercooled by several $10^5\,$K ($\Theta \approx 0.9$).

After 2.95 Gyr of cooling, the central density of the model is $3.65 \times 10^6\,$g/cm$^3$, the temperature becomes lower than $T_m=3.09 \times 10^6\,$K and the core of the crystallizing sequence starts to freeze.  Without additional information about the process of crystallization in a WD, it is difficult to estimate the nucleation rate (and the corresponding degree of undercooling) necessary to initiate macroscopic crystallization on the relevant time scale of order 0.1$\,$Gyr and a mass scale of order $10^{30}\,$g.  Figure \ref{fig:WD_cooling}b shows the delay in crystallization as a function of the required nucleation rate. Based on the parameters of the onset of crystallization in the WD cooling model, we can estimate a conservative lower limit of $J \approx 10^{-25}\,$Gyr$^{-1}\,$cm$^{-3}$, which corresponds to a delay of $0.5\,$Gyr ($\kappa=0.3$). Remarkably, increasing the nucleation rate by 40 orders of magnitude barely increases the delay to $0.7\,$Gyr. In other words, the delay in crystallization due to the undercooling necessary to trigger homogeneous nucleation is about $0.6\,$Gyr and no less than 0.5$\,$Gyr, regardless of assumptions about how crystallization occurs on the spatial and time scales of a white dwarf. Compared to the age of the star ($\sim 3\,$Gyr) and the age of the oldest known white dwarfs ($\sim 10\,$Gyr), this is a significant delay that may be observable as it corresponds to about a 200 to 300K decrease in surface temperature at that time at which crystallization should be expected to begin.

It is important to remember that our calculation of nucleation rates depends on the assumptions of CNT and is only applied for one-component plasmas. This means that there may be other crystallization mechanisms present that would  accelerate nucleation and erase this effect. Further study of nucleation in the multi-component Yukawa plasma will reveal the degree to which nonclassical nucleation modifies these preliminary results. Inhomogeneous or multi-step nucleation may become important, especially in realistic WD plasmas that contain impurities.

We can also apply our cluster size distributions to WD conditions. Fig \ref{fig:clusterDist} shows distributions for supercooled $10^6 \text{g/cm}^3$ carbon plasmas. These distributions suggests that in mildly undercooled one-component WD core plasmas, clusters of $>200$ particles are rare with concentrations of less than $1 \text{ cm}^{-3}$ within 5\% of the melt temperature, however every $\text{cm}^3$ contains millions of clusters with $N_s\approx 100$. Even when nucleation is too slow to cause solidification according to Fig. \ref{fig:WD_cooling}, there is still a large population of transient clusters. Calculating a similar distribution for multi-component WD plasmas could affect our understanding of the rate at which different species are transported through the star \cite{Bauer2020}. 

\section{Summary}

Motivated by open questions in nucleation, including the process of crystallization of WD stars, we have performed brute force and seeded molecular dynamics simulations (with an improved seeded simulation procedure) of nucleation in undercooled YOCP fluids with screening parameters $\kappa = 0,2,\text{ and }5$. We extracted nucleation rates from these simulations and found that for all $\kappa$, those rates become so small for $\Theta\gtrsim 0.9$ that classical homogeneous nucleation cannot explain crystallization in WD within a reasonable time. The simulations indicate that systems with long-range interatomic potentials have slower homogeneous nucleation, larger critical clusters, and fewer pre-critical clusters than similar system with short-ranged potentials. We also used classical nucleation theory to extrapolate our results to near the melt temperature. This CNT extrapolation predicts small concentrations of transient solid clusters of $\sim100$ particles in the bulk supercooled liquid near $T_m$.

Our $\kappa=0$ results are most applicable to WD, but they should be considered preliminary indications of the rates of nucleation and sizes of clusters present in WD because this work focused on homogeneous nucleation of a single-component plasma. The cores of WD and other important systems are typically mixtures of species with different charges and masses. With these caveats in mind, our calculations indicate that the onset of homogeneous nucleation would not occur until 0.6 billion years after it is assumed to begin in typical white dwarf models.

We are extending this work to multi-component Yukawa systems. This will provide more relevant nucleation rates and reveal whether WD mixtures, like C/O/Ne, will contain solid clusters with compositions predicted to drive distillation processes and explain observed WD cooling anomalies \cite{blouinNeDist2021}. It would also allow direct study of whether impurities materially affect nucleation and solidification, perhaps by introducing different crystal structures \cite{kozhberov2021} or seeding inhomogeneous nucleation.

\begin{acknowledgements}
This material is based upon work supported by the Department of Energy [National Nuclear Security Administration] University of Rochester “National Inertial Confinement Program” under Award Number DE-NA0004144.     

The work of JD and DS was performed under the auspices of the US Department of Energy under Contract No. 89233218CNA000001, and was supported by the Laboratory Directed Research and Development program of Los Alamos National Laboratory.  
\end{acknowledgements}

\bibliography{YukawaNucleation}

\appendix

\section{Lennard-Jones Test Case} \label{app:LJ}

\begin{figure}
    \centering
    \includegraphics[width=1\linewidth]{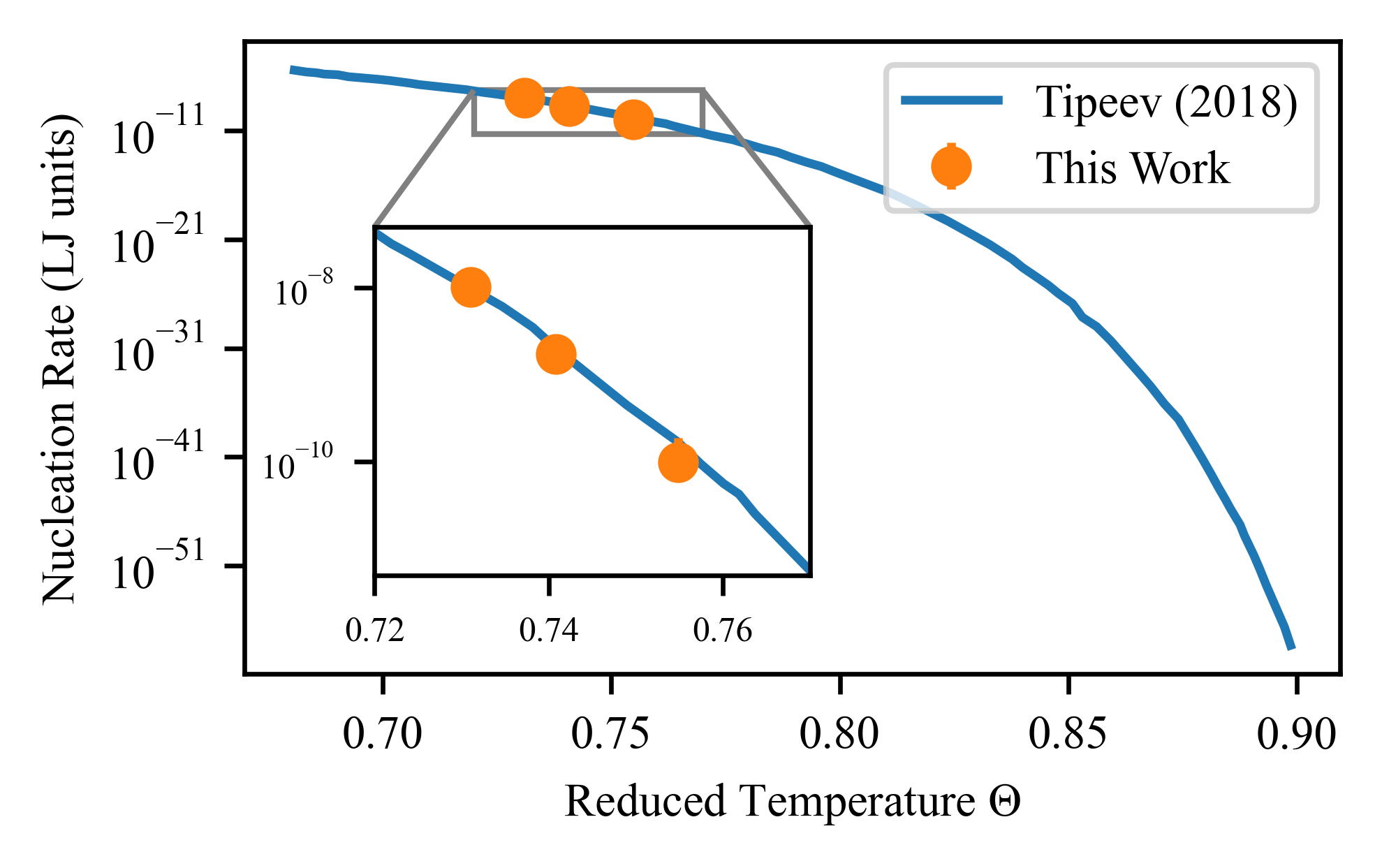}
    \caption{Nucleation rates for the NPT Lennard-Jones system calculated by Tipeev et al. \protect{\cite{seededLJ}} (blue curve) and our MPFT implementation (orange points).}
    \label{fig:LJ_comparison}
\end{figure}

As a validation of our simulation and analysis method, we compare results for the Lennard-Jones system with the simulations of Tipeev et al. in the NPT ensemble \cite{seededLJ}. We replicated the system parameters described in their paper and did brute force simulations at 3 different reduced temperatures, $\Theta= 0.731, 0.741,$ and $0.755$. We analyzed these simulations with the MFPT method described in section \ref{sec:bruteForce}. These results are compared to those of Tipeev et al. in Fig. \ref{fig:LJ_comparison}. They show that our implementation of the MFPT works as expected. 

It is more difficult to compare our seeded simulations to those from Tipeev because we use different methods to determine critical cluster sizes and attachment rates. However, we ran one set of seeded simulations at $\Theta= 0.896$. Our $N^*$ agree within 10\% despite using different cluster size identification methods. Similarly, we found attachment rates within a factor of 2 of those reported by Tipeev, using the averaging procedure described in section \ref{sec:seededSimulation}. 

\section{Size of Seeded Simulations} \label{app:finite_size}

\begin{figure}
    \centering
    \includegraphics[width=1\linewidth]{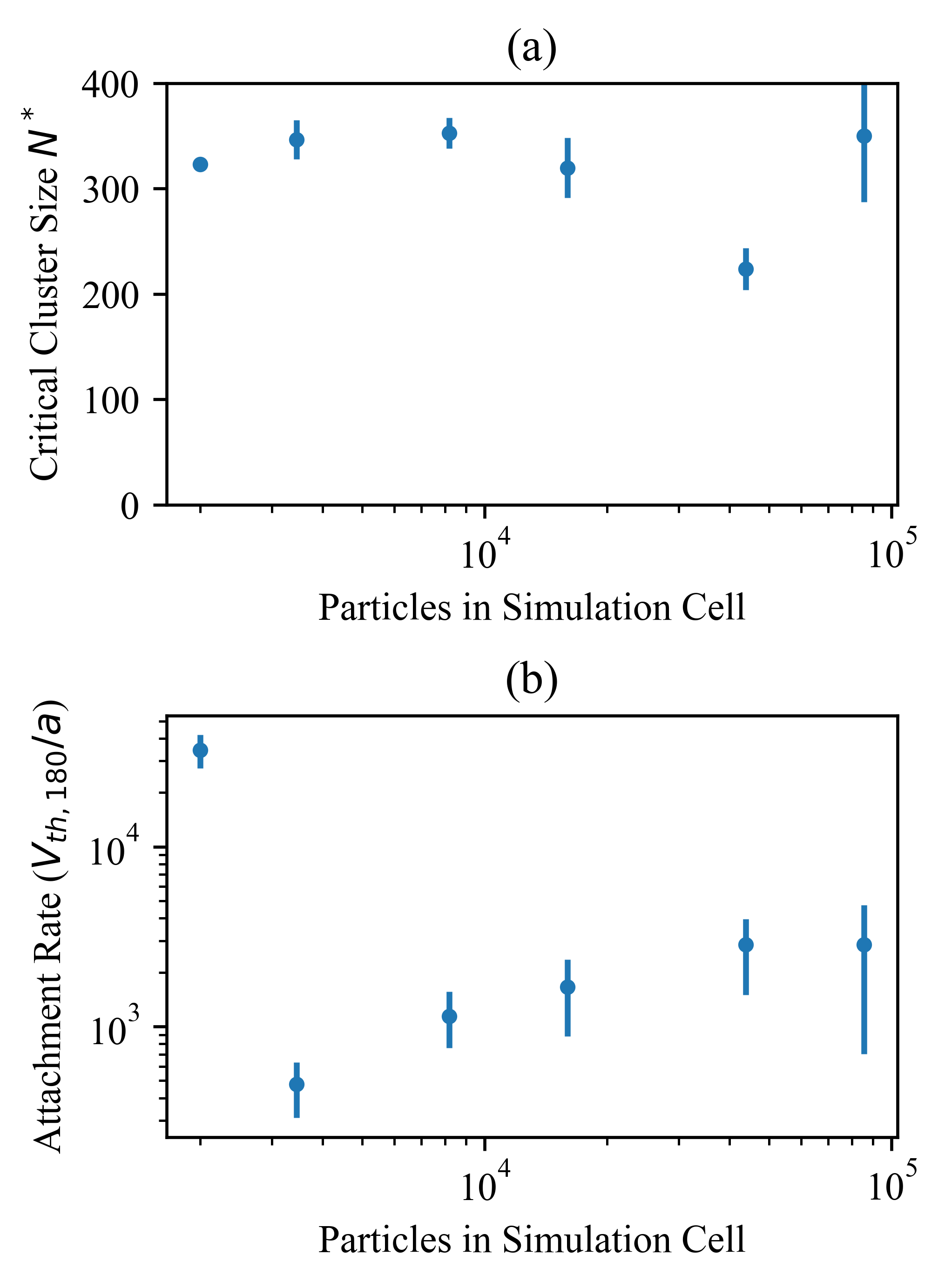}
    \caption{Finite size effects in $\kappa=0$, $\Theta = 0.75$ ($\Gamma = 240$) OCP seeded simulations. Our best estimate for the critical cluster size at these conditions is $N^*=346 \pm 18$ a) Dependence of the critical cluster size $N^*$ on the number of particles in the simulation cell.  b) Dependence of the attachment rate on the number of particles in the simulation cell. For both of these quantities, cells with between $3\times 10^3$ and $10^4$ particles ($10N^*$ to $30N^*$ the critical cluster size) the calculated critical cluster size is independent of simulation size. In larger simulations, the estimates become more poorly constrained due to increased noise.}
    \label{fig:finite_size_test}
\end{figure}

For selected $(\Theta, \kappa)$ conditions, we ran seeded simulations and analysis with different numbers of particles to quantify finite size effects. As an example, Fig. \ref{fig:finite_size_test} shows results for $N^*$ and $D_+$ as a function of the number of particles in the simulation.

These tests  show that the number of particles $N$ in a reliable simulation must satisfy $N>10N^*$. However, the method for identifying cluster sizes introduced in section \ref{sec:seededSimulation} is unreliable for very large simulations. This is because Eqs. \ref{collectiveQ} and \ref{N_decomposition} rely on the collective variable $\overline{Q}_6$, which is a sum over \emph{all} the particles, including the liquid surrounding the cluster. If $N_\ell >> N_s$, random fluctuations in the liquid begin to dominate the value of $\overline{Q}_6$. Therefore it becomes difficult to see the signal of the cluster amidst the fluctuations. This is mitigated when the number of particles in the simulation satisfies $N<30N^*$.

Like $N^*$, calculations of the attachment rate $D_+$ require simulations that are neither too small nor too large. For small cells ($<10 N^*$) the attachment rate is unphysically large because the clusters take up nearly the whole simulation volume and can grow quickly to fill the remaining space. For large cells with $N>30N^*$, the number of detected solid particles, $N_s(t)$, as shown in Fig. \ref{fig:seeded_Array}a becomes noisy due to the large number of liquid particles as discussed above. Because our calculation of the attachment rate [Eq. (\ref{attachment})] relies on the \emph{spread} of this ensemble, noisy  $N_s(t)$ data artificially inflates the rate at which the ensemble appears to spread, increasing the estimated $D_+$.

We find that $N^*$, $D_+$, and $J$ are approximately independent of simulation size when the number of particles is within the $10N^*$ to $30N^*$ range. We believe that this is a good practical recommendation for the size of seeded simulations. All of the results presented in section \ref{sec:seededSimulation} are generated from seeded simulations within this range.

\section{Analytic Fits} \label{app:fits}

Generating the nucleation rate (Fig. \ref{fig:nucRate}) and cluster size distribution (Fig. \ref{fig:clusterDist}) curves required fits to the bulk liquid-solid free energy difference $\Delta \mu$, the free energy associated with the liquid-solid interface $\gamma$, and the rate of attachment of particles onto a critical cluster, $D_+$. these can then be inserted into equations \ref{nucRate} and \ref{distEq}. 

\begin{table}[hbt!]
\setlength{\tabcolsep}{10pt} 
\renewcommand{\arraystretch}{1.5}
    \centering
    \begin{tabular}{c|ccc}
    \hline\hline
          $\kappa$& $A$&$B$&  $D_0$\\ \hline 
          0& 3.441&-1.142&  25.38\\
         2& 3.707& -1.687&13.15\\
         5 (BCC)& 5.025& -3.677&2.981\\
         5 (FCC)& 4.195& -2.693&1.847\\
         \hline \hline
 \end{tabular}
    \caption{}
    \label{tab:fit_params}
\end{table}

We used Farouki and Hamguchi's values for $\Delta \mu$ \cite{Farouki1994, HamaguchiTriplePoint1997}. Our fits to $\gamma$ are of the form \begin{equation*}
    \gamma = A + B\Theta
\end{equation*} where $A$ and $B$ are parameters of the fit and $\gamma$ is in units of $k_BT$. The attachment rate fits are of the form given in Eq. (\ref{eq:attachmentFit}). $D_0$ is the only parameter of the fit and it gives the attachment rate per unit surface area. $D_0$ has units $v_{th,180}/\left(a N_s^{2/3}\right)$ so the time units are defined by the thermal velocity at $\Gamma=180$ and surface area is proportional to $N_s^{2/3}$ assuming a spherical cluster containing $N_s$ particles. The values of the fitting parameters are given in Table \ref{tab:fit_params}.

\end{document}